\documentclass{aa}  

\usepackage{txfonts}
\usepackage{graphicx}

\newcommand{\dd}{\mathrm{d}}

\defcitealias{beck_revised_2005}{BK05}
\newcommand\bk{\citetalias{beck_revised_2005}\xspace}

\usepackage{xpatch}
\usepackage{hyperref}
\hypersetup{
	colorlinks=true,
	breaklinks=true,
	citecolor=blue,
	allcolors=blue,
	frenchlinks=true
}

\makeatletter
\xpatchcmd\NAT@citex
{%
	\@citea\NAT@hyper@{%
		\NAT@nmfmt{\NAT@nm}%
		\hyper@natlinkbreak{\NAT@aysep\NAT@spacechar}{\@citeb\@extra@b@citeb}%
		\NAT@date
	}%
}
{%
	\@citea
	\NAT@nmfmt{\NAT@nm}%
	\NAT@aysep\NAT@spacechar
	\NAT@hyper@{\NAT@date}%
}
{}{}
\xpatchcmd\NAT@citex
{%
	\@citea\NAT@hyper@{%
		\NAT@nmfmt{\NAT@nm}%
		\hyper@natlinkbreak{\NAT@spacechar\NAT@@open\if*#1*\else#1\NAT@spacechar\fi}%
		{\@citeb\@extra@b@citeb}%
		\NAT@date
	}%
}
{
	\@citea
	\NAT@nmfmt{\NAT@nm}%
	\NAT@spacechar\NAT@@open\if*#1*\else#1\NAT@spacechar\fi
	\NAT@hyper@{\NAT@date}%
}
{}{}
\makeatother

\makeatletter
\renewcommand*\aa@pageof{, page \thepage{} of \pageref*{LastPage}}
\makeatother

\begin{document}

   \title{Robust magnetic field estimates in star-forming galaxies with the equipartition formula in the absence of equipartition}

   \author{H.-H. Sandy Chiu\inst{1,2}
          \and
          Mateusz Ruszkowski\inst{2}
          \and
          Maria Werhahn\inst{3}
          \and
          Christoph Pfrommer\inst{1}
          \and
          Timon Thomas\inst{1}
          }

   \institute{Leibniz-Institute for Astrophysics Potsdam (AIP), An der Sternwarte 16, 14482 Potsdam, Germany
         \and
             Department of Astronomy, University of Michigan, Ann Arbor, MI 48109, USA
        \and
             Max-Planck-Institut f{\"u}r Astrophysik (MPA), Karl-Schwarzschild-Str. 1, 85748 Garching, Germany
             }

    \titlerunning{Robust magnetic field estimates with the equipartition formula}
    \authorrunning{Chiu et. al.}

   \date{Received 3 October 2025; accepted 1 June 2026}

 
  \abstract{
   The equipartition model is widely used to estimate magnetic field strength from synchrotron intensity in radio galaxies, yet the validity of its underlying assumptions remains uncertain.
   Using an \textsc{Arepo} simulation which incorporates a two-moment cosmic ray (CR) transport scheme and a multiphase interstellar medium, we compare magnetic fields inferred from synthetic synchrotron emission maps with the true fields in the simulation.
   Starting from the derivation of the equipartition formula, we find that the deviation between the equipartition magnetic field and the true magnetic field depends only weakly on the ratio of the magnetic to the CR energy density. In practice, for both face-on and edge-on projections, the equipartition model slightly overestimates the total synchrotron-weighted magnetic field with mean offsets of 32\% (0.17 dex) and 36\% (0.2 dex), even though the energy equipartition between CR protons and magnetic fields does not hold locally. 
   Beyond these average offsets, a clear trend emerges in edge-on projections that the model underestimates the field in the disk and overestimates it in the halo.
   Our results demonstrate that the validity of the equipartition model depends only weakly on the strict fulfillment of energy equipartition, and that the equipartition model remains a practical method for estimating magnetic field strengths in face-on projection maps based on our CR-magnetohydrodynamics simulation.}


   \maketitle
%

\section{Introduction}\label{sec:intro}

Magnetic fields and cosmic rays (CRs) are fundamental components of the interstellar medium (ISM) in galaxies and play a pivotal role in astrophysics, influencing a broad range of cosmic phenomena \cite[for recent reviews, see][]{zweibel_basis_2017, ruszkowski_cosmic_2023, owen_cosmic_2023}. These components contribute significantly to galaxy evolution and star formation by providing non-thermal pressure support in the ISM \citep{1990ApJ...365..544B, breitschwerdt_galactic_1991, uhlig_galactic_2012, salem_cosmic_2014, kim_introducing_2023, thomas_why_2025, sike_cosmic-ray-driven_2025}, co-regulating star formation \citep{kortgen_global_2019, padovani_impact_2020, martin-alvarez_pandora_2025}, and shaping the distribution of gas across various phases \citep{kortgen_relative_2020, vandevoort_effect_2021}. Magnetic fields and CRs also interact dynamically -- magnetic fields govern CR transport from the ISM into the circumgalactic medium (CGM), thereby affecting the structure of galactic outflows \citep{zweibel_basis_2017, sike_cosmic-ray-driven_2025} and the multiphase distribution of the CGM \citep{weber_crexit_2025}. These outflows carry magnetic flux into the CGM, rearranging the morphology of galactic magnetic fields \citep{pakmor_winds_2016, pakmor_cgm_2020, pfrommer_simulating_2022, kjellgren_dynamical_2025}.

Given the dynamical influence of magnetic fields and CRs on galactic processes, accurately determining the strength and structure of magnetic fields is essential, yet remains observationally challenging for understanding associated non-thermal phenomena. A classical approach involves inferring magnetic field strength from synchrotron emission \citep{pfrommer_estimating_2004, beck_revised_2005} produced by CR electrons gyrating around magnetic field lines \citep{ginzburg_cosmic_1965}. Under the assumptions that (a) the energy densities of CRs and magnetic fields are equal; (b) the CR proton and electron spectra follow a power-law with the same spectral index; and (c) either the vertical ($z$-component) magnetic field is negligible or the field is fully turbulent with an isotropic angular distribution, Beck \& Krause (\citeyear{beck_revised_2005}; hereafter \bk) derived the equipartition formula, which is widely used in observational studies to construct magnetic field strength maps \cite[see \cite{beck_magnetic_2013} for review; ][]{krause_radio_2006, heesen_cosmic_2009, basu_magnetic_2013, miskolczi_chang-es_2019, stein_chang-es_2020, heesen_nearby_2023, nasirzadeh_radiofir_2024, beck_magnetic_2025}.

However, the validity of the widely used equipartition assumption has come under increasing scrutiny. \citet{stepanov_observational_2014} analyzed fluctuations in magnetic fields and synchrotron intensity using observations of the Milky Way and M33, concluding that CR and magnetic energy densities are unlikely to be in equipartition on scales smaller than 100 pc and may, in fact, be slightly anticorrelated. \citet{seta_revisiting_2019} further reinforced this view, demonstrating through test-particle and magnetohydrodynamic (MHD) simulations that equipartition does not hold on scales below the characteristic scale of ISM turbulence. Additionally, \citet{yoast-hull_equipartition_2016}, using combined $\gamma$-ray and synchrotron observations, showed that the equipartition between magnetic fields and CR energy densities breaks down in star-forming regions. Collectively, these studies challenge the robustness of the equipartition assumption, particularly at sub-kiloparsec scales and in dynamically active environments. 

Directly evaluating the validity of the equipartition formula in numerical simulations is challenging, as it requires coupling MHD simulations with CR transport physics and modeling the CR electron spectrum. While recent CR-MHD simulations can evolve the CR spectrum on-the-fly \citep{yang_spatially_2017, girichidis_spectrally_2020, girichidis_spectrally_2022, girichidis_spectrally_2024, ogrodnik_implementation_2021, hopkins_first_2022}, they often rely on simplified transport models, such as constant diffusion coefficients or fixed streaming speeds, despite increasing evidence that CR transport exhibits strong spatial variability \citep{farber_impact_2018, hopkins_testing_2021, armillotta_cosmic-ray_2022, xiao_studying_2025}. On the contrary, the two-moment CR transport schemes, inspired by the analogy between CR and radiative transport, extend the governing equations to include the explicit time evolution of the CR flux, rather than assuming a steady state form \citep{jiang_new_2018, thomas_cosmic-ray_2019, thomas_cosmic-ray-driven_2023, thomas_why_2025}. The two-moment method with spatially varying diffusion coefficient has shown improved agreement with observational signatures by generating CR electron spectrum in post-processing \citep{thomas_probing_2020, chiu_simulating_2024} and represents a key advancement implemented in the galaxy model used in this study compared to other similar studies that also examine the equipartition formula \citep{ponnada_synchrotron_2024, dacunha_overestimation_2024}.

Efforts to evaluate the equipartition formula using MHD simulations have yielded differing conclusions. \citet{ponnada_synchrotron_2024} employed cosmological zoom-in simulations with a spectrally resolved but spatially constant scattering rate CR model and found that the equipartition magnetic field underestimates the synchrotron emission-weighted magnetic field. In contrast, \citet{dacunha_overestimation_2024}, using MHD zoom-in simulations without self-consistently evolved CRs, examined the effects of various CR distribution assumptions and concluded that the equipartition formula tends to overestimate magnetic field strength. Moreover, \citet{linzer_estimation_2025} employed tall-box TIGRESS MHD simulations, post-processed with a two-moment solver that models the transport of spectrally resolved CR protons and electrons. They found that magnetic field estimates based on the equipartition formula agree well with the true magnetic field strength on kiloparsec scales, but the correlation deteriorates on smaller spatial scales. 
These conflicting results highlight the lack of consensus regarding whether the equipartition method systematically over- or underestimates true magnetic field strengths, and to what extent its underlying assumptions influence the outcome. 

In this study, we address these uncertainties by employing CR-MHD simulations that couple the state-of-the-art two-moment CR transport model \citep{thomas_cosmic-ray_2019}, featuring spatially varying diffusion coefficients, with the \textsc{Crisp} ISM model (Thomas et al., in prep.; see also \citealt{thomas_why_2025}). We focus on evaluating whether the assumptions embedded in the equipartition formula lead to significant discrepancies in magnetic field measurements between mock observations and the intrinsic field given in the simulation. The structure of this work is as follows: 
In Section~\ref{sec:method}, we describe the simulation setup, along with the procedures for calculating the CR spectra and the emission map. In Section~\ref{sec:theory}, we present a theoretical derivation, following the approach of the standard equipartition formula, and demonstrate, both analytically and in our simulations, that the ratio between the equipartition magnetic field and the true magnetic field depends only weakly on the ratio between magnetic and CR energy densities. We then assess this theory from an observational perspective by comparing the magnetic field ratio and the energy density ratio in a face-on view (Section~\ref{sec:face-on}) and an edge-on view (Section~\ref{sec:edge-on}). We compare our result with previous literature in Section~\ref{sec:discussion}. Our main findings are summarized in Section~\ref{sec:conclusion}.


\section{Methods} \label{sec:method}

In this section, we describe the simulation setup in detail, followed by the methodology for computing steady-state CR spectra and the resulting synchrotron emissivity. The simulation was originally introduced by \citet{thomas_why_2025} to investigate the differences between thermally and CR-driven galactic winds. Subsequently, \citet{chiu_simulating_2024} generated mock radio observations from the same simulation and demonstrated that the modeled galaxy is consistent with observational radio data in terms of its vertical intensity profile and multi-wavelength spectrum.
\subsection{Simulations}

Our simulations utilized the moving-mesh code \textsc{AREPO} \citep{springel_e_2010, pakmor_improving_2016}, combined with its two-moment cosmic ray magnetohydrodynamics (CRMHD) module \citep{thomas_cosmic-ray_2019, thomas_finite_2021, thomas_comparing_2022}, and the ISM model \textsc{Crisp} (Cosmic Rays and InterStellar Physics), which captured key ISM processes and cosmic ray interactions (Thomas et al., in prep.; see also \citealt{thomas_why_2025}). We begin by detailing the simulation setup, followed by an overview of the two-moment CR transport method and the \textsc{Crisp} framework.

We simulated an isolated Milky Way-like galaxy hosted in a static Hernquist halo of mass $M_{\rm 200}=10^{12}~\mathrm{M}_\odot$, the concentration parameter $c_{\rm halo}=7$, and baryon mass fraction $f_\mathrm{b}=0.155$ \citep{hernquist_analytical_1990}, initialized with doubly exponential profiles for both the stellar and gaseous disks. The gaseous disk had a radial scale length of 5~kpc, a scale height of 0.5~kpc, and a total mass of $8\times 10^9~\mathrm{M}_\odot$, while the stellar disk shared the same geometry and had a total mass of $3.2\times 10^{10}~\mathrm{M}_\odot$. In this initial condition setup, star particles were assigned a mass of $M_\star=1000~\mathrm{M}_\odot$ and distributed randomly following the exponential disk profile. Gas cells were initialized analogously and undergo adaptive (de-)refinement to maintain mass and volume within a factor of 2 (or no less than 0.5) of $1000~\mathrm{M}_\odot$ and $V = (4\pi/3)r^3$, respectively, ensuring a resolution of $r = 100$~pc or finer -- roughly several times smaller than the typical observational beam size\footnote{As a reference, a recent study that adopts the equipartition formula has a resolution of 340~pc \citep{beck_magnetic_2025}.}. 

Star formation was modeled by converting gas cells with a number density above $100~\mathrm{cm}^{-3}$ into stars on the local free-fall timescale, which is 4.5~Myr at the threshold density. Stellar feedback was implemented using outputs from \texttt{STARBURST99} \citep{leitherer_starburst99_1999}, which prescribe the return of mass, metals, and energy to the ISM. Each supernova contributed $1.06\times 10^{51}$~erg of mechanical energy, with 5\% allocated to CR injection. The initial magnetic field consisted of a toroidal component, $B_{\rm tor}=10^{-1}\sqrt{\rho/\rho_{\rm max}}~\mu\mathrm{G}$, where $\rho$ is the gas density that has a maximum $\rho_{\rm max}$ at the galaxy center, and a uniform vertical component of $10^{-3}~\mu$G. These seed fields were quickly amplified by small-scale dynamo processes, leading to a final magnetic field configuration that is insensitive to the initial conditions.

The CR energy injected by supernovae then evolved according to the two-moment CRMHD scheme \citep{thomas_cosmic-ray_2019}. This framework tracked both the CR energy density ($\varepsilon_{\rm CR}$) and flux density using a gray approximation, wherein the momentum-dependent distribution was not resolved explicitly, and energy loss rates were averaged over the CR momentum spectrum. The CR transport velocities were determined by interactions with small-scale Alfv\'en waves, which are subject to damping processes such as non-linear Landau damping \citep{miller_magnetohydrodynamic_1991} and ion-neutral damping \citep{kulsrud_effect_1969}. Beyond wave-particle interactions, CRs also lose energy through hadronic and Coulomb interactions \citep{pfrommer_simulating_2017}.

Since CR transport was regulated by ion-neutral damping, which is highly sensitive to the thermochemical state of the ISM, we employed the \textsc{Crisp} module (Thomas et al., in prep.) to model the multi-phase ISM environment. \textsc{Crisp} self-consistently evolved the ionization states of 12 species, including $\rm H_2$, \ion{H}{I}, \ion{H}{II}, all ionization stages of He, and the first two ionization stages of C, O, and Si. The module accounted for heating from far-UV radiation emitted by young stars, along with a range of cooling processes: Ly$\alpha$ emission from hydrogen, rotational-vibrational transitions of $\rm H_2$ \citep{moseley_turbulent_2021}, bremsstrahlung cooling at high temperatures \citep{cen_hydrodynamic_1992}, and metal line cooling. The latter was calculated either directly from collisional rates \citep{abrahamsson_fine-structure_2007, grassi_krome_2014} for $T < 10^4$~K, or interpolated from precomputed \textsc{Chianti} tables \citep{dere_chianti_1997} for $T > 10^4$~K.

\subsection{Modeling CR spectra}

We model CR spectra with the \textsc{CRAYON+} code, following the procedure outlined in \citet{werhahn_cosmic_2021}, with modifications described in \citet{chiu_simulating_2024} to incorporate the effects of spatially varying CR transport as captured by the two-moment framework. We adopted a cell-based steady-state approximation, assuming that the CR energy density evolved on timescales longer than the cooling timescales of CR protons and electrons, such that the source and loss terms in the diffusion-loss equation were in equilibrium. This assumption was supported by \citet{werhahn_cosmic_2021} and further validated in the appendix of \citet{chiu_simulating_2024}. Under this assumption, we solved the diffusion-loss equation separately for protons, primary electrons, and secondary electrons in each cell.

The CR injection spectrum ($q_i$) for each species was modeled as a power-law in momentum with an exponential cutoff:
\begin{equation}
    q_i(p_i)\,\mathrm{d}p_i = C_i\,p_i^{-\alpha_{\rm inj}}\,\exp\left[-\left(p_i/p_{{\rm cut},i}\right)^n\right]\,\mathrm{d}p_i,
    \label{eq:injection spectrum}
\end{equation}
where $p_i = P_i/(m_ic) = \sqrt{[E_i/(m_ic^2)]^2 - 1}$ is the normalized momentum, $E_i$ is the total particle energy, $m_i$ is the particle rest mass, and $c$ is the speed of light. The subscript $i$ denotes the CR species ($i = \mathrm{p}, \mathrm{e}$ for protons and electrons, respectively). The normalization factor $C_i$ is set through a subsequent renormalization procedure, as described below. $\alpha_{\rm inj}$ is the injection spectral index and is set to 2.1 for both protons and primary electrons \citep{werhahn_cosmic_2021-2}. $p_{\rm cut, i}$ is the cut-off momentum that is different for protons \citep[$p_{\rm cut, p}=1\ {\rm PeV}/m_{\rm p}c^2$, ][]{gaisser_cosmic_1990} and electrons \citep[$p_{\rm cut, e}=20\ {\rm TeV}/m_{\rm e}c^2$, ][]{vink_supernova_2011}. We set $n=1$ for protons and $n=2$ for primary electrons \citep{zirakashvili_analytical_2007, blasi_shock_2010}.

The loss term in the diffusion-loss equation consisted of cooling and escape losses. For cooling losses, we included hadronic and Coulomb losses for CR protons, as well as bremsstrahlung, synchrotron, and inverse Compton (IC) losses for primary electrons \citep[see details in sections 3.1.2 and 3.1.3 in ][]{werhahn_cosmic_2021}. For escape losses, we included particle escape due to advection, diffusion, and streaming losses. We considered an energy-dependent diffusion coefficient $D(E)=D_{\rm 0, eff}(E/E_0)^{0.3}$ \citep{werhahn_cosmic_2021-1}, where $E_0$ = 3~GeV is a free parameter. We emulated the CR self-confinement picture with a dominant CR streaming process and define the effective diffusion coefficient $D_{\rm 0, eff}$ as $|\varv_{\rm CR}| L_{\rm CR}$, where $\varv_{\rm CR}=3f_{\rm CR}/4\varepsilon_{\rm CR}$, $f_{\rm CR}$ is the CR proton energy flux, $\varepsilon_{\rm CR}$ is the CR proton energy density, and $L_{\rm CR}=\varepsilon_{\rm CR}/|\nabla\varepsilon_{\rm CR}|$ is the diffusion length in each cell \citep{chiu_simulating_2024}. The advection timescale was calculated by $\tau_{\rm adv}=L_{\rm CR}/\varv_z$, where $\varv_z$ is the $z$-direction velocity. 

The CR spectra resulting from the diffusion-loss equation were then renormalized so that in each cell, the total kinetic energy density of the CR protons $\int^\infty_0 T_{\rm p}(p_{\rm p})f_{\rm p}(p_{\rm p})\dd p_{\rm p}$ equals the CR energy density in our CRMHD model. Here, $T_{\rm p}(p_{\rm p})/(m_{\rm p}c^2)=\sqrt{p_{\rm p}^2+1}-1$ is the kinetic energy of each momentum bin and $f_{\rm p}$ is the CR proton spectral energy density. The primary CR electron spectra were normalized with the electron-to-proton injection number ratio $K^{\rm inj}_{\rm ep}$ at a fixed normalization kinetic energy $T_{\rm norm}=$10~GeV. Specifically,
\begin{equation}
    K^{\rm inj}_{\rm ep}(T_{\rm norm}) \cdot q_{\rm p}(p_{\rm norm,p})\dd p_{\rm p}  = q^{\rm prim}_{\rm e}(p_{\rm norm, e}) \dd p_{\rm e}
    \label{eq:kinj}
\end{equation}
where $q_{\rm p}$ and $q^{\rm prim}_{\rm e}$ are the injection spectrum for protons and primary electrons as defined in Equation~\eqref{eq:injection spectrum}, and $p_{\rm norm, e/p}=\sqrt{[T_{\rm norm}/(m_{\rm e/p} c^2)+1]^2 - 1}$. The parameter $K^{\rm inj}_{\rm ep}$ can be approximately related to the electron-to-proton energy injection fraction
\begin{equation}
    \zeta_{\rm prim}=\frac{\varepsilon^{\rm inj}_{\rm e}}{\varepsilon^{\rm inj}_{\rm p}}=\frac{\int^\infty_0 T_{\rm e}q_{\rm e}\dd p_{\rm e}}{\int^\infty_0 T_{\rm p}q_{\rm p}\dd p_{\rm p}}
\end{equation}
by 
\begin{equation}
    K^{\rm inj}_{\rm ep}=\zeta_{\rm prim, no-cutoff}\left(\frac{m_{\rm p}}{m_{\rm e}}\right)^{2-\alpha_{\rm inj}}
    \label{eq:kep}
\end{equation}
as long as the exponential cutoff in the injection spectrum can be neglected. For an injection slope of $\alpha_{\rm inj} = 2.1$, we found that omitting the cutoff introduces a $\sim$25\% error in the parameter $\zeta_{\rm prim}$, such that $\zeta_{\rm prim, no-cutoff} = 1.25\,\zeta_{\rm prim}$ \citep{chiu_simulating_2024}. One-zone models suggested that $\zeta_{\rm prim}$ typically falls in the range of 10--20\% \citep{lacki_physics_2010}, corresponding to an electron-to-proton injection ratio of $K^{\rm inj}_{\rm ep} = 0.059$--$0.12$. Based on our prior analysis in \citet{chiu_simulating_2024}, we adopted $K^{\rm inj}_{\rm ep} = 0.092$ in this study.

\subsection{Radio emission}

We calculated the synchrotron emission following \citet{RybickiLightman} as 
\begin{equation}
    j_\nu=\frac{\sqrt{3}e^3B_\perp}{m_{\rm e} c^2}\int^{\infty}_0f_{\rm e}(p_{\rm e})F(\nu/\nu_c)\dd p_{\rm e}, 
    \label{eq:syn_j}
\end{equation}
where $e$ denotes the elementary charge, $B_\perp$ is the component of the magnetic field perpendicular to the line of sight, and $f_{\rm e} = \dd N_{\rm e}/(\dd E_{\rm e}\dd V)$ represents the CR electron spectral density (i.e., the number of electrons per unit energy and unit volume). The observed frequency is denoted by $\nu$, and the synchrotron emission kernel is given by $F(x) \equiv x\int_x^\infty K_{5/3}(\zeta)\,\dd \zeta$, where $K_{5/3}(\zeta)$ is the modified Bessel function of order 5/3. We adopted the analytical approximation to $F(\nu/\nu_{\rm c})$ provided by \citet{aharonian_angular_2010}. We integrated the synchrotron emissivity along the line of sight with the radiative transfer equation, neglecting the free-free and synchrotron self-absorption since the corresponding optical depth at the considered frequency of 1.4~GHz is much smaller than unity. 


\begin{figure*}
    \centering
    \includegraphics[width=\linewidth]{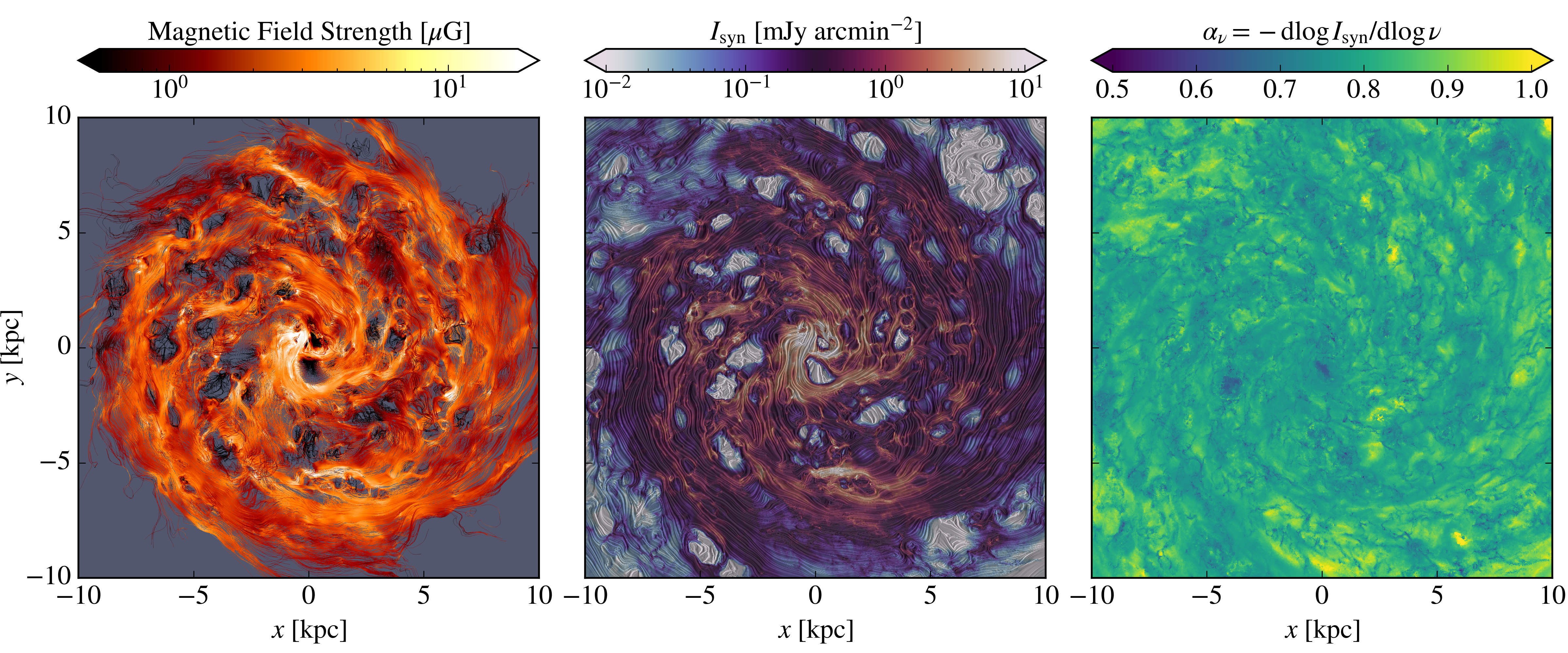}
    \caption{Left: Magnetic field lines color-coded by magnetic field strength. Middle: Synchrotron intensity shown in the background, with magnetic field directions derived from the Stokes Q and U parameters overlaid in the foreground. The effect of Faraday rotation is not included. Right: Radio spectral slope map calculated between 140 MHz and 1.4 GHz. The spectral index steepens from approximately $\alpha_\nu = 0.55$ (corresponding to $\alpha_\mathrm{e} = 2.1$, our $\alpha_{\rm inj}$) in the spiral arm to $\alpha_\nu = 0.9$ (corresponding to $\alpha_\mathrm{e} = 2.8$) in the inter-arm region. The middle and right panel are calculated with projected synchrotron intensity using a depth of 1~kpc along the line of sight.}
    \label{fig:map}
\end{figure*}

\section{Results} \label{sec:result}

Figure~\ref{fig:map} shows a face-on view of our simulated galaxy. The left panel shows magnetic field lines traced through the galactic disk. We start the magnetic field line tracing in the galactic midplane ($z=0$) on a grid of $128\times64$ points that are equally spaced in radius and polar angle out to $R=9~\mathrm{kpc}$. We stop following a magnetic field line once it moves into the CGM ($|z| > 0.5~\mathrm{kpc}$) or outside the shown frame. The tracing process naturally highlights regions with a high magnetic field strengths, while few magnetic field lines pass through regions with a low magnetic field strength, such as feedback-driven bubbles. These bubbles appear to be enshrouded by magnetic field lines that pass by the bubble interior and map out their rims. The middle panel presents the corresponding synchrotron intensity map, overlaid with magnetic field orientations inferred from the Stokes $Q$ and $U$ parameters \citep[for details on the calculation of Stokes parameters, see Section 2.4 of][]{chiu_simulating_2024}. As illustrated in the figure, the synchrotron intensity traces both the morphology and strength of the magnetic field. In particular, bubble-like features appear at the same locations in both maps, and regions with stronger magnetic fields exhibit higher synchrotron intensity. However, the intrinsic magnetic field exhibits a significantly more detailed structure than the orientation inferred from observations, emphasizing the need to evaluate how accurately synchrotron intensity traces the underlying magnetic field distribution. 

In this section, we use synthetic observations to evaluate the validity of the equipartition formula. We first show that the accuracy of the equipartition formula depends only weakly on the assumption of equipartition at the level of cell-by-cell comparisons\footnote{For the resolution setup in our simulation, the scale is around 100 pc. See also discussion in Appendix \ref{apx:scale}.}. We then examine how the correlation changes under projection effects when applying different weighting variables, considering both face-on and edge-on views.

\subsection{What decides the accuracy of the equipartition formula?}
\label{sec:theory}
The magnetic field strength is commonly determined from the observed synchrotron intensity using the equipartition formula in \bk:
\begin{equation}
B_{\text{eq}} = \left\{ \frac{4\pi (2\alpha_\nu + 1)(K_0 + 1)\, I_\nu\, E_{\text{p}}^{1 - 2\alpha_\nu} \left( \nu / 2c_1 \right)^{\alpha_\nu}}{\left[(2\alpha_\nu - 1)\, c_2(\alpha_\nu)\, l\, c_4\right]} \right\}^{1/(\alpha_\nu + 3)}
\label{eq:equipartition}
\end{equation}
where $B_{\rm eq}$ is the equipartition magnetic field, $\alpha_\nu$ is the observed radio spectral index that relates with the CR electron spectral index as $\alpha_\nu=(\alpha_\mathrm{e}-1)/2$, $I_\nu$ is the synchrotron intensity, $\nu$ is the observed frequency, $K_0$ is the proton-to-electron number density ratio, $E_{\rm p}=938.28~$MeV is the proton rest energy, $c_1$ and $c_2$ are combinations of physical constants while $c_2$ depends on $\alpha_\nu$, $l$ is the length of integration along the line of sight, and $c_4$ is the converting factor between the total magnetic field and the field component projected on the sky ($B_{x-y}$ for simulation viewing face-on). 

\begin{figure}
    \centering
    \includegraphics[width=1\linewidth]{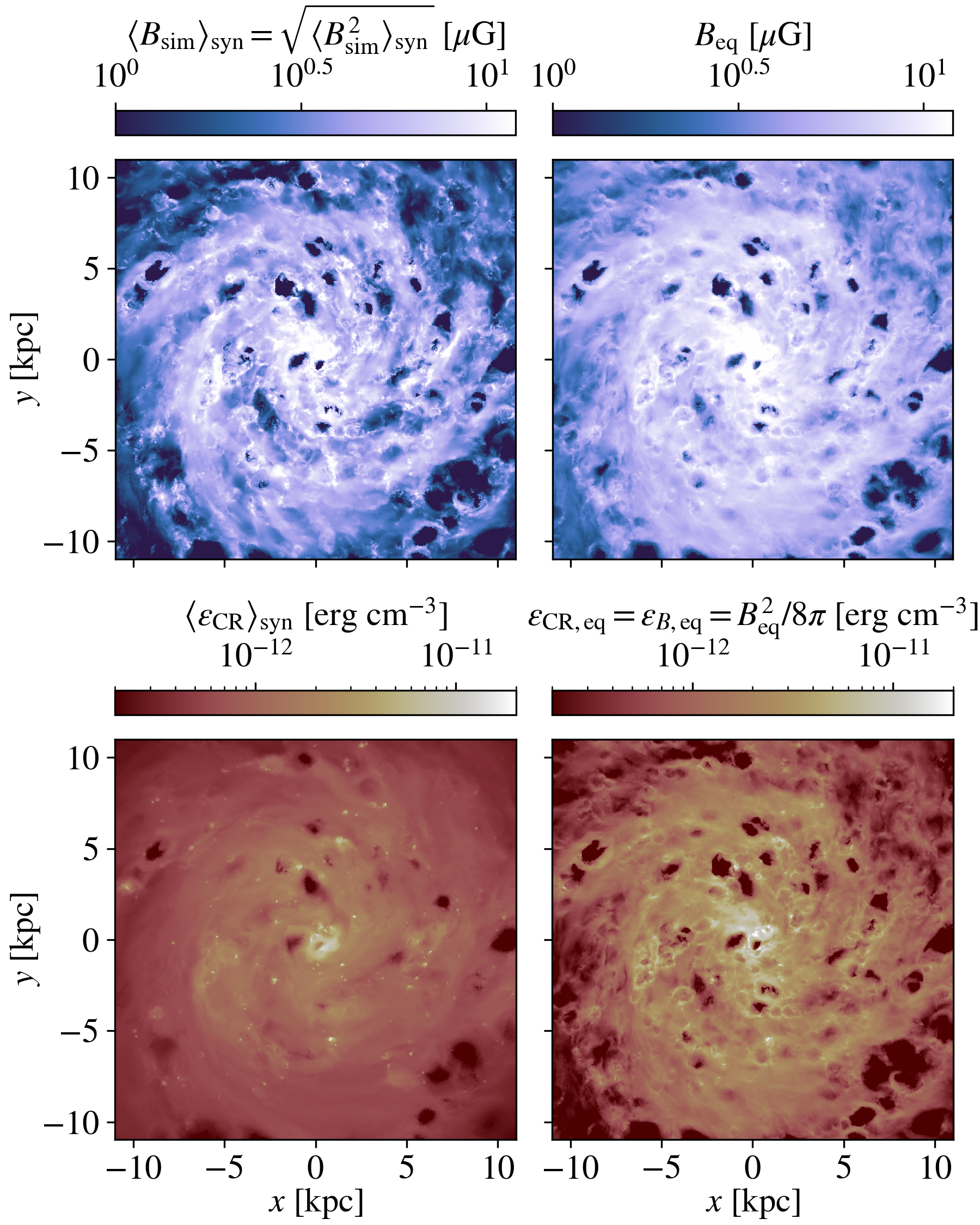}
    \caption{Comparison of the true magnetic field strength from the simulation (upper left) with the equipartition magnetic field (upper right), and the true CR energy density (lower left) with the equipartition energy density (lower right). Both the true magnetic field and the CR energy density are weighted by synchrotron emission. The equipartition magnetic field is derived by inserting the synchrotron intensity and radio spectral index into Equation~\eqref{eq:equipartition}, assuming $K_0 = 100$, $l = 1$, and $c_4 = 1$. All panels show projections integrated over $\pm 0.5$ kpc above and below the disk ($|z|$<0.5~kpc). This comparison demonstrates that while the equipartition magnetic field can partially recover the true magnetic field morphology (due to the weak dependence of the field ratio on the energy density ratio), the CR energy density calculated from the equipartition magnetic field fails to accurately trace the CR energy density.}
    \label{fig:eb_vs_ecr}
\end{figure}

In the top-right panel of Fig.~\ref{fig:eb_vs_ecr}, we present the equipartition magnetic field from our simulation. The equipartition magnetic field $B_{\rm eq}$ is calculated using the synchrotron intensity at 1.4\,GHz (middle panel of Fig.~\ref{fig:map}) and the observed spectral index $\alpha_\nu$ between 1.4 and 0.14\,GHz (right panel of Fig.~\ref{fig:map}). The result is insensitive to the choice of the frequency used to calculate $\alpha_\nu$. Observationally-motivated values are adopted for the remaining free parameters in the formula\footnote{
The parameter $K_0$ differs from $K^{\rm inj}_{\rm ep}$. The quantity $K^{\rm inj}_{\rm ep}$ describes the electron-to-proton ratio of the injected spectrum, whereas $K_0$ characterizes the proton-to-electron ratio of the observed (cooled) spectrum. Because electrons cool faster than protons, an effect that becomes increasingly important at higher energies, the electron-to-proton ratio in the cooled (observed) spectrum is reduced relative to the injected spectrum. In our simulations, the mean electron-to-proton ratio in the cooled spectrum is $0.01$, consistent with $K_0 = 1 / K^{\rm obs}_{\rm ep} = 100$, where $K^{\rm obs}_{\rm ep}$ denotes the electron-to-proton ratio of the observed (cooled) spectrum.
}
: $l = 1$~kpc, $K_0 = 100$, and $c_4 = 1$, corresponding to a completely face-on view of the galaxy and assuming no $z$-component in the magnetic field.

We find that the magnetic field values (maps in the top row) exhibit a higher degree of similarity to each other than the energy density values (maps in the bottom row). In particular, the equipartition magnetic field (top-right panel of Fig.~\ref{fig:eb_vs_ecr}) reproduces the main morphological features of the true magnetic field (top-left panel of Fig. \ref{fig:eb_vs_ecr})\footnote{In this paper, the weighted magnetic field is always calculated by the square root of the synchrotron luminosity- or volume-weighted magnetic field squared, i.e., $\langle B_{\rm sim}\rangle_{\rm syn/vol}=\sqrt{\langle B^2\rangle_{\rm syn/vol}}$, along the line of sight.}: regions where the equipartition field exhibits depressions generally coincide with similar features in the true magnetic field map. In contrast, the equipartition CR energy density (bottom-right) displays pronounced holes that are not present in the true CR energy density map, which remains comparatively smooth.

\begin{figure}
    \centering
    \includegraphics[width=\linewidth]{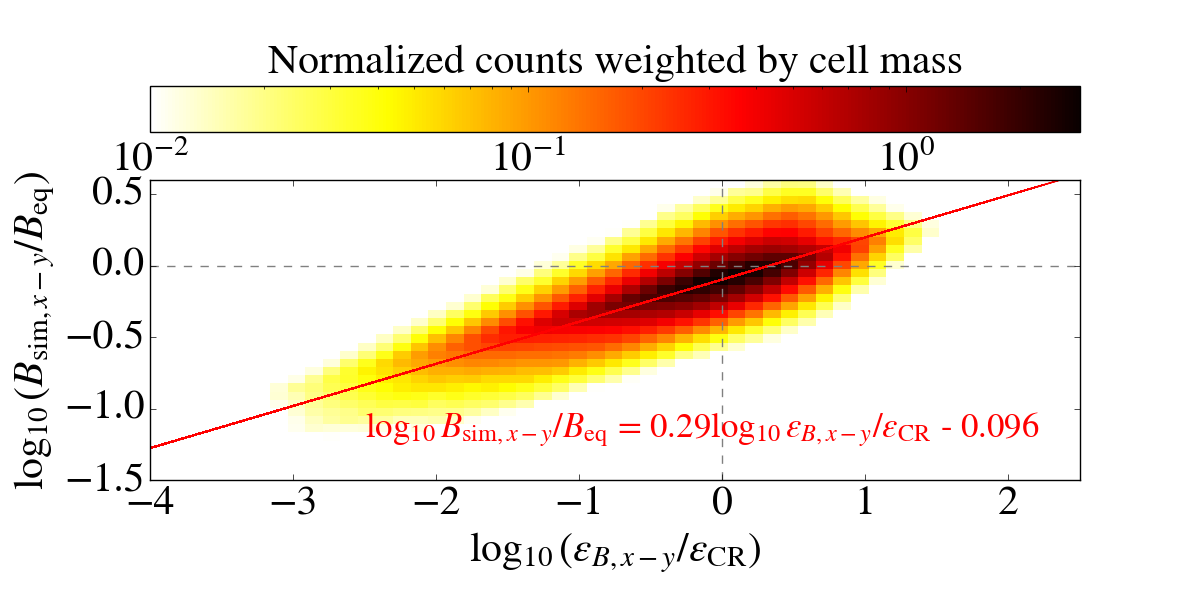}
    \caption{Correlation between the magnetic field ratio and the energy density ratio in individual computational cells within the disk ($|x| < 10$ kpc, $|y| < 10$ kpc, and $|z| < 2.5$ kpc). The magnetic field strength is computed using the field components in the plane of the galaxy. The equipartition magnetic field in each cell is calculated using the synchrotron intensity, $I_{\rm syn}=L_{\rm syn}/4\pi (\pi r_{\rm cell}^2)$, where $r_{\rm cell}$ is the radius of each gas cell, and the corresponding $\alpha_\nu$, assuming $K_0 = 100$, $c_4 = 1$, and $l = r_{\rm cell}$. The best-fit line across the cells is shown in red, with a slope of 0.29 and an intercept of $-0.096$. The gray-dash line shows where $B_{{\rm sim}, x-y}=B_{\rm eq}$ and $\varepsilon_{B,x-y}=\varepsilon_{\rm CR}$. This result is consistent with our analytical derivation and confirms that the magnetic field ratio depends only weakly on the energy density ratio. 
    }
    \label{fig:cell_by_cell_comparison}
\end{figure}

This suggests that the equipartition magnetic field can partially recover the synchrotron-weighted magnetic field, while the galaxy may not be in equipartition. The correlation between the magnetic field ratio ($B_{\rm sim}/B_{\rm eq}$) and the energy density ratio ($\varepsilon_B/\varepsilon_{\rm CR}$) can be analytically derived as we will show in the following:

First, the CR energy density depends on both the synchrotron intensity and the magnetic field strength as
\begin{align}
    \varepsilon_{\rm CR} \propto \frac{I_\nu}{B^{\alpha_\nu+1}}
    \label{eq: I_nu}
\end{align}
where we have assumed that $B_\perp=c_4B$. For a full derivation including all constants, see Appendix~\ref{app:derivation}.
Assuming equipartition $\varepsilon_\mathrm{CR}=\varepsilon_{B, {\rm eq}}=B_{\rm eq}^2/(8\pi)$, we have
\begin{align}
    I_\nu \propto B_{\rm eq}^{\alpha_\nu +3}.
    \label{eq:I_nu_Beq}
\end{align}
In the simulation, we can write down the ratio of CR to magnetic energy density as
\begin{align}
    \frac{\varepsilon_\mathrm{CR, sim}}{\varepsilon_{B, \mathrm{sim}}}\propto \frac{I_\nu }{B_\mathrm{sim}^{\alpha_\nu+3}}.
    \label{eq:epsilon_CRsim/epsilon_Bsim}
\end{align}
Combining equations \eqref{eq:I_nu_Beq} and \eqref{eq:epsilon_CRsim/epsilon_Bsim} leads to
\begin{align}
\frac{\varepsilon_\mathrm{CR, sim}}{\varepsilon_{B, \mathrm{sim}}}\propto \left(\frac{B_{\rm eq}}{B_\mathrm{sim}}\right)^{\alpha_\nu +3}
\end{align}
or, equivalently, 
\begin{align}
    \frac{B_{\rm eq}}{B_\mathrm{sim}} \propto \left(\frac{\varepsilon_\mathrm{CR, sim}}{\varepsilon_{B, \mathrm{sim}}}\right)^{\frac{1}{\alpha_\nu +3}}.
    \label{eq:proportional_correlation}
\end{align}
Taking into account all constants, the exact relation (see Appendix~\ref{app:derivation}) reads 
\begin{align}
        \log\frac{B_{\rm sim}}{B_{\rm eq}} =&\frac{1}{\alpha_\nu+3}\log\left(\frac{\varepsilon_{B, {\rm sim}}}{\varepsilon_{\rm CR, sim}}\right)\nonumber \\
    &-\frac{\alpha_\nu+1}{\alpha_\nu+3}\log\left(\frac{c_{\rm 4,sim}}{c_{\rm 4, eq}}\right)+\frac{1}{\alpha_\nu+3}\log\left(\frac{C_{\rm sim}}{C_{\rm eq}}\right).
    \label{eq:log-relation}
\end{align}
where $C_{\rm sim}$ and $C_{\rm eq}$ are the proportional constant in Equation~\eqref{eq: I_nu}. From the proportional relation in Equation~\eqref{eq:proportional_correlation}, we can see that the scaling between the magnetic field ratio and the energy density ratio is weak: for a typical spectral index $\alpha_\nu \sim 0.5-0.8$, the exponent is only $\sim 0.29-0.26$. This means that even if the ratio of CR to magnetic energy density is wrong by an order of magnitude, the derived equipartition magnetic field only changes by less than a factor of two.

To test the above derivation, we consider the correlation between the magnetic field ratio and the energy density ratio at the level of each computational cells in Fig.~\ref{fig:cell_by_cell_comparison}. We simplify the test by neglecting all the magnetic component in the $z$-direction by calculating the magnetic energy density as $B_{x-y}^2/8\pi$, so that the $c_{\rm 4, sim}/c_{\rm 4,eq}$ term becomes 1. We assume that gas cells are spherical with radius $r_{\rm cell}=(3V/4\pi)^{1/3}$. The synchrotron intensity is calculated as 
\begin{equation}
    I_{\rm syn}=\frac{L_{\rm syn}}{4\pi (\pi r_{\rm cell}^2)}
\end{equation}
where $L_{\rm syn}$ is the synchrotron luminosity in each cell, $\pi r_{\rm cell}^2$ represents the surface area, and $4\pi$ is to consider the solid angle assuming the emission is isotropic. 

The resulting correlation between the field ratio and the energy density ratio is shown in Fig.~\ref{fig:cell_by_cell_comparison}. The fitted slope in Fig.~\ref{fig:cell_by_cell_comparison} is very close to an injected CR spectral index $\alpha_\mathrm{e}=2.1$, $\alpha_\nu=0.55$, and the slope $1/(\alpha_\nu+3)=0.28$. For the synchrotron intensity-weighted average spectral index we find from the simulation, $\alpha_\nu=0.7$, the logarithmic slope in Equation~\eqref{eq:log-relation} becomes 0.27, which is also close to what we obtained in Fig.~\ref{fig:cell_by_cell_comparison}. The shallow slope indicates that the equipartition magnetic field scales weakly with the ratio between the CR and magnetic energy densities.
For example, if the equipartition assumption is violated such that the CR energy density is ten times larger than the magnetic energy density, the equipartition magnetic field would overestimate the true field strength by only a factor of $10^{0.29} \approx 1.9$. 

The scatter in the data points arises from variations in $\alpha_\nu$ across different cells, as well as potential deviations of the simulated CR energy density from the values predicted by Equation~\eqref{eq: I_nu}, which can also explain the small offset of 0.08 (see Appendix~\ref{apx:ecr_ratio}). This offset of $-0.096$ implies a slight overestimation of the equipartition magnetic field relative to the true magnetic field when the system is in exact equipartition (i.e., $B_{{\rm sim}, x-y} = 0.8B_{\rm eq}$ when $\varepsilon_B = \varepsilon_{\rm CR}$). We note that the magnitude and sign of this offset may vary among different simulations and viewing inclinations, and therefore it should not be interpreted as a general tendency for the equipartition formula to systematically overestimate or underestimate the true magnetic field.

Free parameters such as $K_0$, $l$, and the inclination factor $c_4$ in Equation~\eqref{eq:equipartition} are also influenced by the $1/(\alpha_\nu + 3)$ exponent. For example, even if the integrated length along the line of sight $l$ becomes 0.1 instead of 1~kpc due to the small volume filling factor as suggested by \citet{ponnada_synchrotron_2024}, $B_{\rm eq}$ will only change by $10^{0.26-0.29}=1.82$--$1.95$, which is less than a factor of 2.

\subsection{Face-on projection maps}\label{sec:face-on}

\begin{figure*}
    \centering
    \includegraphics[width=\linewidth]{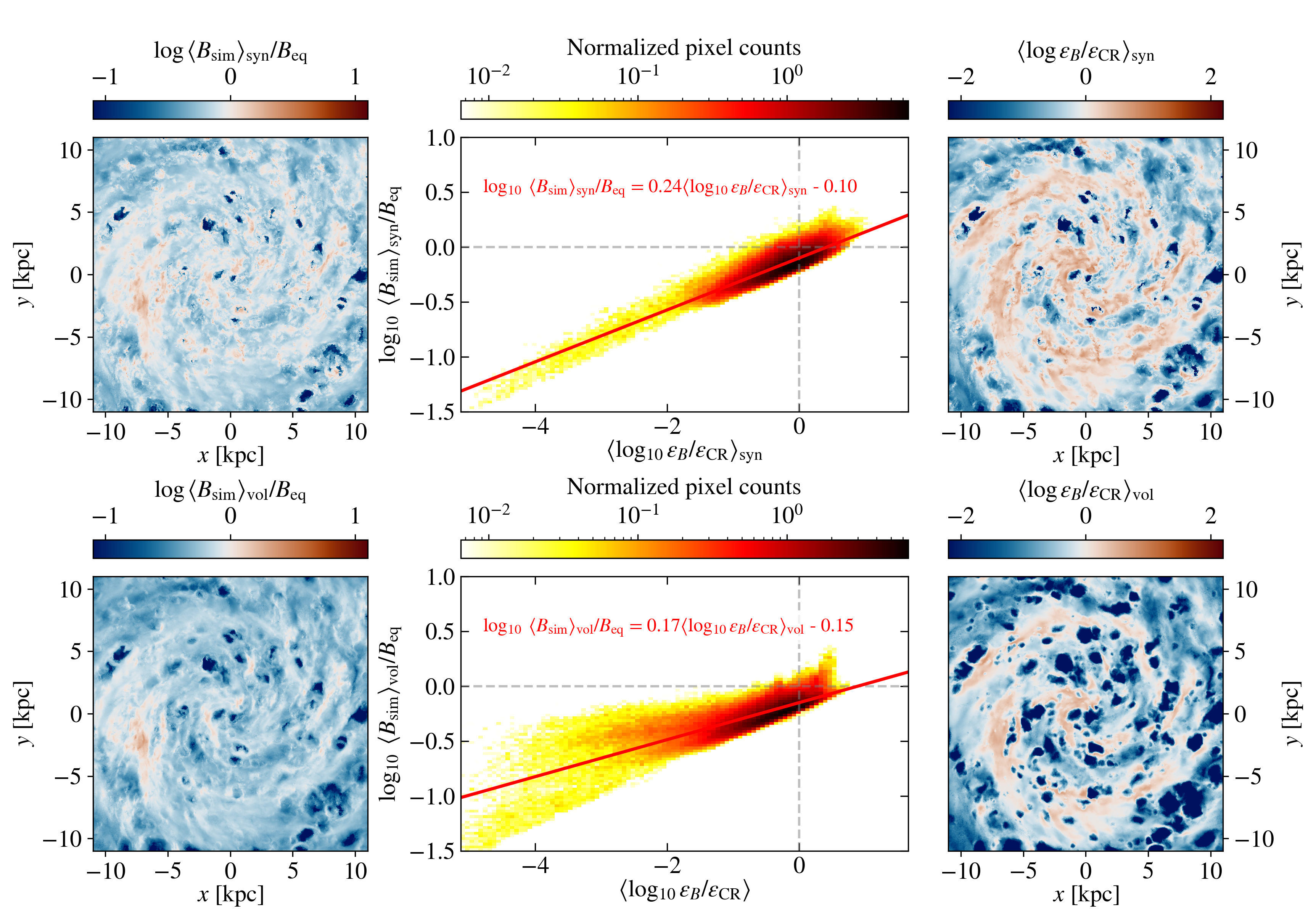}
    \caption{Left column: Ratio of the magnetic field in the simulation weighted by synchrotron emissivity ($\langle B_{\rm sim}\rangle_{\rm syn}$) to the field inferred from mock observation based on the equipartition assumption \citep[$B_{\rm eq}$;][]{beck_revised_2005}, assuming $K_0=100$, $c_4=1$, and $l=1$~kpc. Right column: Synchrotron-weighted average of the log of the ratio of magnetic to CR energy density. Middle column: Pixel-by-pixel correlation between the corresponding maps in the left and right columns, with best-fit lines overlaid. The upper panels show results weighted by synchrotron emission, while the lower panels show results weighted by volume. Weighting by volume rather than synchrotron emission reduces both the slope and the intercept of the correlation. All maps are computed using a projection depth of 1 kpc along the line of sight.}
    \label{fig:map_faceon}
\end{figure*}
In this section, we test the equipartition formula using the intensity map and projected quantities with different weighting schemes to mimic real observations. The equipartition magnetic field is calculated with commonly adopted observational parameters, including $K_0 = 100$ and $l = 1$~kpc. For a direct comparison with observational data, both the simulated magnetic field strength $B_{\rm sim}$ and the magnetic energy density $\varepsilon_B$ are evaluated using the total magnetic field, without excluding the $z$-component (see Appendix~\ref{apx:offset} for discussion about the $z$-component magnetic field), in contrast to the approach taken in the previous section.

In the left column of Fig.~\ref{fig:map_faceon}, we show the ratio between the simulated magnetic field weighted by synchrotron emission or volume ($\left<B_{\rm sim}\right>_{\rm syn/vol}$) and the equipartition magnetic field inferred from the mock synchrotron emission ($B_{\rm eq}$). The ratio of these two magnetic fields indicates whether the \bk formula recovers the ``true'' magnetic field. The right column shows the average logarithmic ratio between the magnetic and CR energy density\footnote{Taking the logarithmic ratio before averaging along the line of sight makes the result independent of how the ratio is defined, i.e., $\left<\log\varepsilon_B/\varepsilon_{\rm CR}\right>=-\left<\log\varepsilon_{\rm CR}/\varepsilon_{B}\right>$. If the energy density ratio is averaged before taking the logarithm, then the symmetry no longer holds, i.e., $\log\left<\varepsilon_B/\varepsilon_{\rm CR}\right> \neq -\log\left<\varepsilon_{\rm CR}/\varepsilon_B\right>$. See also Appendix A in \cite{chiu_simulating_2024}.} weighted by synchrotron emission or volume, quantifying to what extent the equipartition assumption is justified. 

We find that our equipartition magnetic field can successfully recover the synchrotron-weighted magnetic field as the mean value of the $\log_{\rm 10}\langle B_{\rm sim}\rangle_{\rm syn}/B_{\rm eq}$ peaks around -0.17 dex, meaning that the equipartition magnetic field slightly overestimates the synchrotron-weighted magnetic field as $\langle B_{\rm sim}\rangle_{\rm syn}\approx B_{\rm eq}\cdot 0.69$. 

The deviation of the equipartition estimate from the true magnetic field, quantified by $\log(B_{\rm sim}/B_{\rm eq})$, depends approximately linearly, but weakly, on $\log(\varepsilon_B/\varepsilon_{\rm CR})$. This behavior is consistent with the analytic relation derived in Section~\ref{sec:theory}.
The positive correlation between the field ratio and the energy density ratio can be understood as the equipartition assumption effectively inferring the magnetic energy density from the CR energy density. When the magnetic energy density exceeds the CR energy density (shown in red in the right panel), assuming equipartition leads to an underestimation of the magnetic field, resulting in a larger $\langle B_{\rm sim}\rangle_{\rm syn}/B_{\rm eq}$ (also shown in red in the left panel), and vice versa.
In Figure~\ref{fig:map_faceon}, the colorbars in the right panels span twice the logarithmic range of those in the left panels to account for the fact that the energy density scales with the square of the magnetic field. We find that the right panels appear more clumpy than the left panels, indicating that the spatial fluctuations in the energy density ratio are significantly more extreme than those in the magnetic field ratio. This indicates that the deviations of $B_{\rm eq}$ from $\langle B_{\rm sim} \rangle_{\rm syn}$ are much smaller than the deviations introduced by the equipartition assumption itself.

In the middle panel of Fig.~\ref{fig:map_faceon}, we present the pixel-by-pixel correlation between the magnetic field ratio (left panel) and the energy density ratio (right panel), with the best-fit line shown in red. In the upper panel, where the synchrotron-weighted quantities are compared, the slope is 0.24, corresponding to $\alpha_\nu = 1.17$ and $\alpha_\mathrm{e} = 3.3$, which is a reasonable value for a cooled CR electron spectral index. This result differs slightly from the cell-by-cell comparison shown in Fig.~\ref{fig:cell_by_cell_comparison} because, in the case of projections, synchrotron-bright cells dominate along the line of sight. This leads to a steeper spectrum compared to the cell-by-cell analysis, where no projection was applied prior to the analysis and each cell was weighted equally. Nevertheless, the slope remains small, indicating that the equipartition field can still recover the magnetic field even when the energy densities are not in equipartition. The non-zero offset in the fitted line is likely due to projection effects and the simplified assumption about the correlation among $\varepsilon_{\rm CR}$, $I_\nu$ and $B_{\perp}$. Detailed discussion about the offset term can be found in Appendix~\ref{apx:offset}.

So far, we have compared the equipartition magnetic field only with synchrotron-weighted properties, since $B_{\rm eq}$ is positively correlated with synchrotron intensity and thus effectively traces the synchrotron weighted magnetic field, $\langle B_{\rm sim}\rangle_{\rm syn}$. However, the volume-weighted magnetic field, $\langle B_{\rm sim}\rangle_{\rm vol}$, is more physically meaningful for estimating the magnetic pressure.

In the lower panel of Fig.~\ref{fig:map_faceon}, we find that the equipartition magnetic field overestimate the volume-weighted magnetic field with a mean value of $\log_{\rm 10}\langle B_{\rm sim}\rangle_{\rm vol}/B_{\rm eq}$ peaks around -0.27, showing that, as in the case of the synchrotron-weighted magnetic field, the equipartition magnetic field significantly overestimates the volume-weighted magnetic field. This is because $\langle B_{\rm sim}\rangle_{\rm vol}$ gives more weight to diffuse, volume-filling gas that may contribute weakly to synchrotron emission. As a result, $\langle B_{\rm sim}\rangle_{\rm vol}$ is systematically lower than $\langle B_{\rm sim}\rangle_{\rm syn}$, consistent with previous findings \citep{ponnada_synchrotron_2024, whittingham_zooming-cluster_2024}.

\begin{figure}
    \centering
    \includegraphics[width=\linewidth]{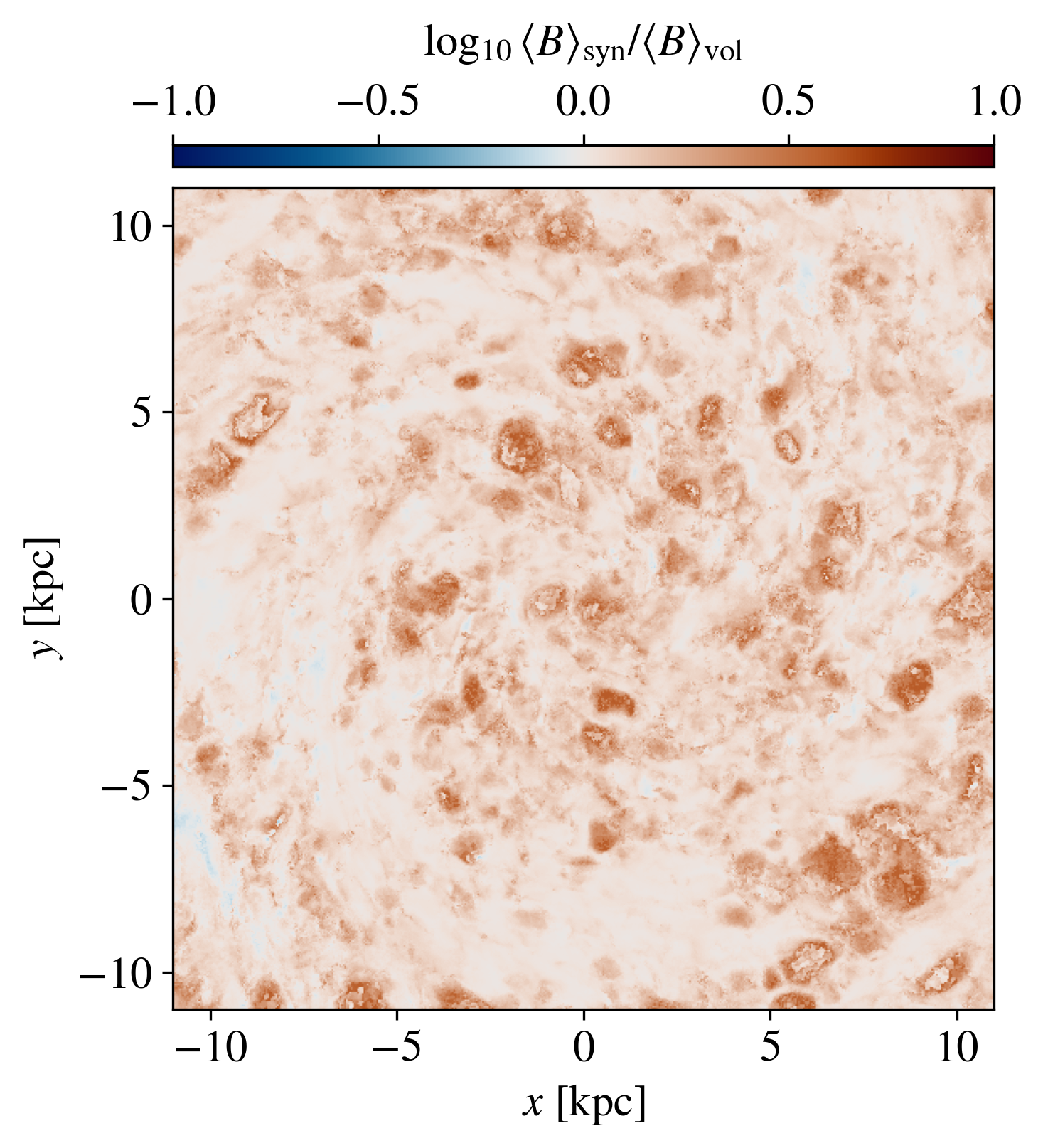}
    \caption{ Ratio of the synchrotron-weighted average $B$-field ($\langle B_{\rm sim}\rangle_{\rm syn}$) and the volume-weighted average $B$-field ($\langle B_{\rm sim}\rangle_{\rm vol}$). $\langle B_{\rm sim}\rangle_{\rm syn}$ is in general larger than $\langle B_{\rm sim}\rangle_{\rm vol}$, and is much larger than $B_{\rm vol}$ around the star-forming region. The ratio involving $\langle B_{\rm sim}\rangle_{\rm vol}$ is computed using a projection depth of 1~kpc along the line of sight. Both $B_{\rm syn}$ and $\langle B_{\rm sim}\rangle_{\rm vol}$ are therefore evaluated within the same 1~kpc line-of-sight region, without additional spatial cuts. This limited projection depth effectively selects regions within the galactic disk and avoids bias from the more diffuse, volume-filling halo.
    }
    \label{fig:syn_or_vol}
\end{figure}

Figure~\ref{fig:syn_or_vol} further supports this conclusion by showing the spatial distribution of the ratio of $\langle B_{\rm sim}\rangle_{\rm syn}$ and $\langle B_{\rm sim}\rangle_{\rm vol}$. The map reveals that $\langle B_{\rm sim}\rangle_{\rm syn}$ is generally larger than its volume-weighted counterpart, particularly at the outer edges of star-forming regions.
The reason is that gas cells in these regions tend to have stronger magnetic fields but occupy smaller volumes. In conclusion, the synchrotron-weighted field emphasizes these high-field, low-volume regions, while the volume-weighted field emphasizes the volume-filling lower density ISM which has a lower synchrotron emissivity and magnetic field strengths.

The lower panel of Fig.~\ref{fig:map_faceon} also shows that the dependence of the field ratio on the energy density ratio is weaker (with a smaller slope of the fitting line) when we switch the weighting to volume. As discussed in Section~\ref{sec:theory}, the slope reflects the collective behavior of individual pixels, each of which may exhibit a different spectral index, $\alpha_\nu$. Therefore, as weighting by synchrotron emission emphasizes dense, spiral arm regions where CRs are freshly injected, volume-weighting highlights diffuse, inter-arm regions where CRs have cooled as they diffuse outward from the spiral arms. This spatial variation in the radio spectral index is also evident in the right panel of  Fig.~\ref{fig:map}, where the inter-arm and outer regions of the galaxy display higher values of $\alpha_\nu$. As $\alpha_\nu$ increases, the slope of the fitted relation naturally decreases (see Eq.\ref{eq:log-relation}) when the fit is performed using volume weighting.

\subsection{Edge-on projection maps}\label{sec:edge-on}

\begin{figure*}
    \centering
    \includegraphics[width=1\linewidth]{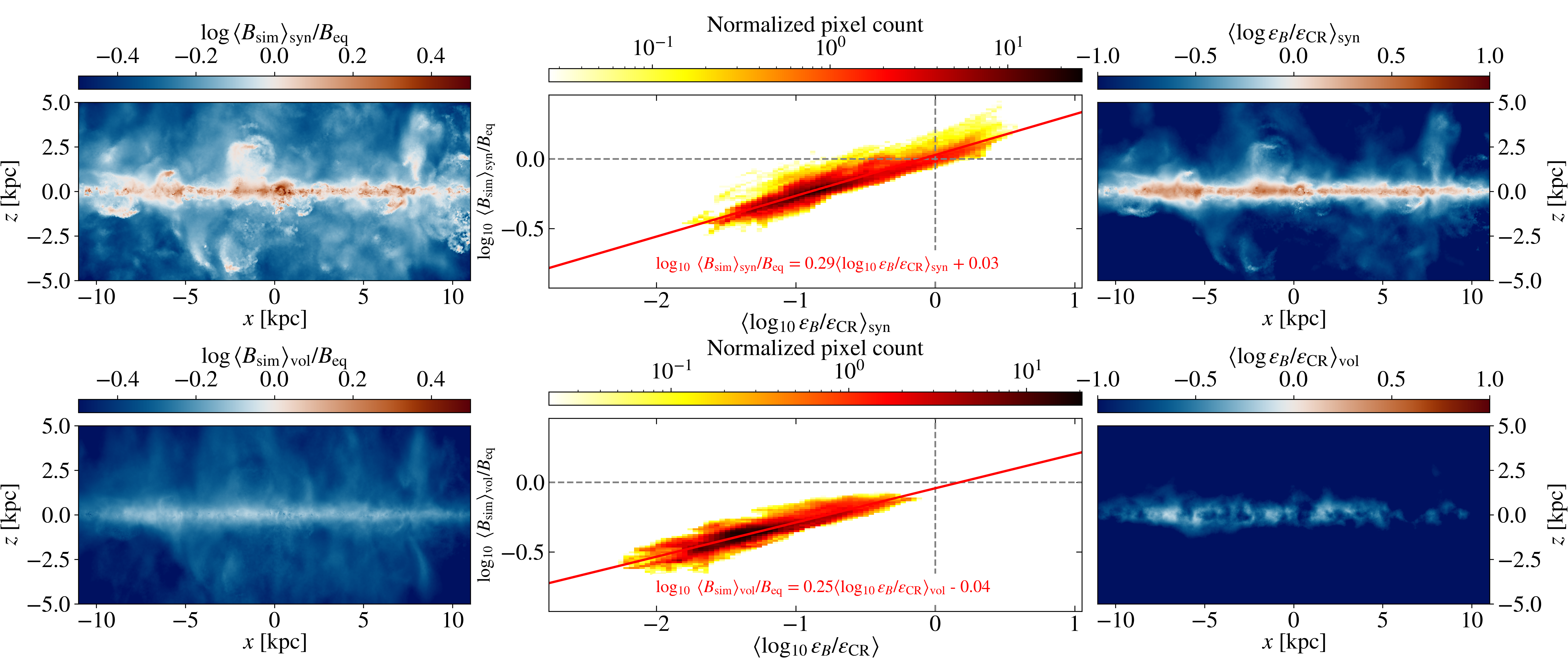}
    \caption{Same as Fig.~\ref{fig:map_faceon}, but shown edge-on. All panels are computed with a projection depth of 22 kpc along the line of sight.
    Since the magnetic energy density exceeds the CR energy density in the disk, the equipartition method tends to overestimate the synchrotron-weighted magnetic field in the halo and underestimate it in the disk.
    }
    \label{fig:map_edgeon}
\end{figure*}
We examine the equipartition formula from the edge-on view in Fig.~\ref{fig:map_edgeon}. Again, we adopt typical values for the free parameters that observers would use when applying the \bk formula to an edge-on galaxy, including $K_0 = 100$, $c_4 = (2/3)^{(\alpha_\mathrm{e} + 1)/4}$, assuming a fully turbulent magnetic field with an isotropic angle distribution, and $l = 2R\sqrt{1 - (x/R)^2 - (z/Z)^2}$, where $X=14.1$~kpc and $Z=8.3$~kpc are the semi-major and semi-minor axes, respectively, determined by fitting an ellipse to the intensity contour above 18\% of the peak intensity \citep{schmidt_chang-es_2019}. 

In the edge-on view, the equipartition formula underestimates the synchrotron-weighted magnetic field in the disk and overestimates it in the halo ($|z| > 1$~kpc). This arises from the magnetic energy density exceeding the CR energy density in the disk, and the opposite in the halo. Due to the positive correlation between the energy density ratio and the magnetic field ratio, the equipartition field systematically deviates from the true field in opposite directions in these regions.

Similar behavior of the magnetic field ratio and the slope of the fitting line is observed when the weighting is switched to volume. In the lower panel of Fig.~\ref{fig:map_edgeon}, the equipartition magnetic field overestimates the volume-weighted magnetic field in the halo and nearly recovers the true field in the disk. The mean of the ratio $\log_{10}(\langle B_{\rm sim}\rangle_{\rm vol}/B_{\rm eq})$ in each pixel peaks around -0.37, reflecting the fact that $\langle B_{\rm sim}\rangle_{\rm syn}>\langle B_{\rm sim}\rangle_{\rm vol}$ in general. The slope of the fitting line likewise decreases when the magnetic field and energy density are weighted by volume rather than synchrotron emission, for the same reason discussed in relation to Fig.~\ref{fig:map_faceon}.

The intercept, or offset, of the fitting line (i.e., the last term on the right-hand side in either of the fitting formulae in the middle panels in Fig.~\ref{fig:map_faceon} and Fig.~\ref{fig:map_edgeon}) decreases as the weighting shifts from synchrotron weighting to volume weighting. This offset corresponds to the intercept of the relation between $B_{\rm eq}$ and $B_{\rm sim}$ evaluated at exact equipartition (i.e., when $\log(\varepsilon_B/\varepsilon_{\rm CR}) = 0$). It therefore represents the systematic bias of the equipartition estimate relative to the true magnetic field at the equipartition point.
For instance, in the fitting line shown in Fig.~\ref{fig:map_edgeon}, when $\log \varepsilon_B/\varepsilon_{\rm CR} = 0$, we have $\langle B_{\rm sim} \rangle_{\rm syn} = B_{\rm eq} \cdot 10^{0.03}$, which is greater than $\langle B_{\rm sim} \rangle_{\rm vol} = B_{\rm eq} \cdot 10^{-0.04}$. Thus, the decrease in the intercept is a natural consequence of $\langle B_{\rm sim} \rangle_{\rm syn}$ being larger than $\langle B_{\rm sim} \rangle_{\rm vol}$. 

\section{Comparison to other work} \label{sec:discussion}

Other studies employing various numerical approaches report results consistent with ours, namely that the over- or underestimation of the magnetic field by the equipartition formula depends on the underlying magnetic field strength.

\citet{ponnada_synchrotron_2024} use cosmological zoom-in FIRE simulations with a spectrally resolved CR transport model to generate synthetic synchrotron emission maps and assess the equipartition formula. They argue that synchrotron emission is predominantly produced in cool, dense gas. As a result, the fiducial equipartition model underestimates the true magnetic field by a factor of 2-3 when the galaxy is viewed face-on.

\citet{dacunha_overestimation_2024} analyze galaxies formed in cosmological zoom-in MHD simulations without explicit CR transport and examine how well the equipartition formula recovers the intrinsic magnetic field under different assumptions about the CR distribution, including a uniform CR distribution and CR energy density equal to the magnetic energy density. They find that, although the equipartition formula generally overestimates the intrinsic magnetic field by a factor of 10-50, the degree of overestimation depends on the true magnetic field strength. 

\citet{linzer_estimation_2025} use TIGRESS tall-box MHD simulations and obtain CR spectra via post-processing with a two-moment CR transport model. On kiloparsec scales, they find that the equipartition magnetic field is broadly consistent with the true magnetic field, with $B_{\rm eq}$ exceeding the intrinsic field by a factor of approximately 1.4. On smaller scales, they report\footnote{We found $B_{\rm eq}\propto B_{\rm sim}^{0.57}$ with our data. See discussion in Appendix~\ref{apx:scale}.} $B_{\rm eq} \propto B_{\rm sim}^{0.36}$, implying that the equipartition field overestimates the true field when the intrinsic field is weak and underestimates it when the intrinsic field is strong. This trend is consistent with the findings of \citet{dacunha_overestimation_2024}.

The over- or underestimation of the true magnetic field strength by the equipartition magnetic field estimate can be understood from our derivation in Equation~\eqref{eq:equipartition}. When the magnetic field is weak, the ratio of CR to magnetic energy density is large, making $B_{\rm eq}$ more likely to exceed $B_{\rm sim}$. This framework also explains the apparent differences among these studies. The typical magnetic field strength in the disk is approximately 10~$\mu$G in \citet{ponnada_synchrotron_2024}, 4~$\mu$G in \citet{linzer_estimation_2025}, and 4~$\mu$G in \citet{dacunha_overestimation_2024}, while our simulation yields an average field strength of 3~$\mu$G. The magnetic field in \citet{ponnada_synchrotron_2024} is nearly a factor of two stronger than in the other studies. Consequently, \citet{linzer_estimation_2025}, \citet{dacunha_overestimation_2024}, and our work find that the equipartition field tends to overestimate the intrinsic field, whereas \citet{ponnada_synchrotron_2024} report the opposite behavior. This interpretation is consistent with the discussion in both \citet{dacunha_overestimation_2024} and \citet{linzer_estimation_2025}, who show that the ratio between the equipartition and intrinsic magnetic fields depends on the magnetic field strength itself.

Given that $B_{\rm eq} < B_{\rm sim}$ in regions of strong magnetic field (and vice versa), the equipartition magnetic field effectively traces a smoothed representation of the magnetic structure within the galactic disk. In spiral arms and star-forming regions, where the intrinsic magnetic field is strong, $B_{\rm eq}$ tends to underestimate the true field, implying that the actual magnetic field strength exceeds that inferred from equipartition. For an edge-on view of the galaxy, this suggests that the true magnetic pressure gradient is likely stronger than that derived from the equipartition estimate.

The degree of overestimation reported in \citet{dacunha_overestimation_2024} (a factor of 10–50) is substantially larger than that found in \citet{linzer_estimation_2025} (a factor of 1.4) and in our study (a factor of 1.3), despite similar magnetic field strengths. This discrepancy arises from the CR distribution assumed in \citet{dacunha_overestimation_2024}. In their model, the CR number density profile is assumed to be similar to that of the Milky Way. As shown in their figure~5, this CR number density, and therefore the CR energy density, is larger than the equipartition CR number density implied by the magnetic field strength in their simulation.
If instead a similar approach to that adopted in \citet{linzer_estimation_2025} were used, namely assuming a uniform CR energy density in approximate equipartition with the magnetic energy density on large scales, the equipartition formula would likely yield magnetic field estimates closer to the intrinsic values. Indeed, \citet{dacunha_overestimation_2024} note that adopting higher CR electron densities results in a smaller overestimation factor of approximately 2–3, which would be more consistent with the results of \citet{linzer_estimation_2025} and our study.

\section{Conclusions} \label{sec:conclusion}

In this work, we use CR-MHD simulations of an isolated Milky Way-like galaxy to assess the validity of the classical equipartition formula for estimating magnetic field strengths. 
By generating synthetic synchrotron intensity maps and applying the equipartition formula, we compare the inferred magnetic field with the synchrotron-weighted magnetic field from the simulation. 
We find that, despite the equipartition assumption not holding, the equipartition magnetic field closely traces the synchrotron-weighted field in face-on inclinations. For edge-on inclinations, the equipartition magnetic field underestimates the true field strength in the disk and overestimates it in the halo. This reflects the fact that the magnetic energy density exceeds the CR energy density in the disk.

Specifically, we compare the correlation between the ratio of the equipartition magnetic field and the synchrotron-weighted magnetic field (the ``field'' ratio) versus the ratio between CR and magnetic energy density (the ``CR-to-magnetic energy density'' ratio) using both face-on and edge-on projection maps.
We find that for both inclinations, the deviation of the intrinsic magnetic field from the equipartition field depends only weakly on the ratio of magnetic to cosmic ray energy densities, following a shallow power-law relation. This is consistent with analytical expectations determined by the exponent in the equipartition formula.

The equipartition magnetic field traces the synchrotron-weighted magnetic field more accurately than the volume-weighted field. Since the volume-weighted magnetic field is systematically lower, the equipartition field tends to overestimate both the field strength and the corresponding magnetic pressure. The dependence of the field ratio on the energy density ratio is weaker when switching to volume weighting, as this emphasizes the volume-filling regions where CRs have cooled after propagating from their birth sites into more diffuse environments, leading to a larger spectral index and a reduced slope.

Based on the above findings, we conclude that the equipartition magnetic field is a useful tool for estimating magnetic field strength, even though CR energy density may deviate from the magnetic energy density. However, important caveats remain when using the equipartition magnetic field to infer the magnetic pressure or even the CR pressure.
Firstly, we conclude that the deviation between $B_{\rm sim}$ and $B_{\rm eq}$ depends weakly on the energy density ratios and the free parameters used in the equipartition formula based on Equation~\eqref{eq:log-relation}, where all terms in the right hand side contribute to the deviation between the true field and the equipartition field. However, the magnetic pressure is proportional to the magnetic field square, meaning that when we compare $\log (\varepsilon_{B, \rm sim}/\varepsilon_{B, \rm eq})=\log( B_{\rm sim}^2/B_{\rm eq}^2)$, all the terms in the right hand side of Equation~\eqref{eq:log-relation} will be multiplied by 2 and lead to a stronger dependence. 
Secondly, the equipartition magnetic pressure ($B_{\rm eq}^2/8\pi$) should be interpreted only as an upper limit of the true magnetic pressure. This is because the equipartition field more closely traces the synchrotron-weighted magnetic field, which tends to overestimate the volume-weighted field, the more appropriate measure for calculating magnetic pressure.
Finally, the fact that $B_{\rm eq}$ closely matches the magnetic field in simulations does not imply that energy equipartition between CRs and magnetic fields holds, and the resulting magnetic field pressure should not be used to infer the CR pressure. Overall, while the equipartition formula remains a practical and observationally accessible tool, its limitations must be carefully considered when drawing physical inferences beyond field strength estimates.

\begin{acknowledgements}
M.R. and S.C. thank Lucia Armillotta and Sam Ponnada for stimulating discussions, and the Leibniz-Institut f\"ur Astrophysik Potsdam (AIP) for its hospitality during the visit in which part of this work was finalized.
M.R. acknowledges support from the National Science Foundation Collaborative Research Grant NSF AST-2009227.
This work was supported in part through computational resources and services provided by Advanced Research Computing (ARC) -- a division of Information and Technology Services (ITS) at the University of Michigan, Ann Arbor.
TT and CP acknowledge support by the European Research Council under ERC-AdG grant PICOGAL-101019746. The authors gratefully acknowledge the computing time granted by the Resource Allocation Board and provided on the supercomputer Emmy/Grete at NHR-Nord@G\"ottingen as part of the NHR infrastructure. The calculations for this research were conducted with computing resources under the project bbp00070.

\end{acknowledgements}

\section*{Data availability}
The data underlying this article will be shared on reasonable request to the corresponding author.
\bibliography{example}{}
\bibliographystyle{aa}

\begin{appendix}
\section{Weak dependence of energy density ratio and the field ratio}\label{app:derivation}
The correlation between the magnetic field ratio ($B_{\rm sim}/B_{\rm eq}$) and the energy density ratio ($\varepsilon_{B}/\varepsilon_{\rm CR}$) can be quantified following a similar derivation in \bk. Starting with Equation~\eqref{eq: I_nu}:
\begin{equation}
    \varepsilon_{\rm CR} = C\frac{I_\nu}{B_{\perp}^{\alpha_\nu+1}},
    \label{eq:ecr}
\end{equation}
where $C=f(\alpha_\mathrm{e})\frac{K_{0}+1}{l}$ is the proportionality constant that is not a function of the magnetic field, $f(\alpha_\mathrm{e})$ is a constant that only depends on the CR electron spectral index $\alpha_\mathrm{e}$, and $B_{\perp}$ is the magnetic field component perpendicular to the line of sight. To obtain an expression for the equipartition magnetic field, we assume $\varepsilon_{\rm CR, eq}=\varepsilon_{B, {\rm eq}}$ and get
\begin{equation}
    \varepsilon_{\rm CR, eq}=C_{\rm eq}\frac{I_\nu}{B_{\perp, \rm eq}^{\alpha_\nu+1}}=\varepsilon_{B, {\rm eq}}=\frac{B_{\rm eq}^2}{8\pi},
    \label{eq:ecr_eq}
\end{equation}
where $C_{\rm eq}$ is the proportionality constant with observationally motivated $K_0$ and $l$. $\varepsilon_{\rm CR,eq}$ is the CR proton energy density derived from \bk, which is different from the CR proton energy density in simulations ($\varepsilon_{\rm CR, sim}$).
Following \bk, we assume $B_\perp=c_4 B$, and solve for equipartition magnetic field from Equation~\eqref{eq:ecr_eq}:
\begin{equation}
       B_{\rm eq}^{\alpha_\nu+3}=8\pi C_{\rm eq} I_\nu c_{\rm 4,eq}^{-\alpha_\nu-1}.
       \label{eq:B_eq}
\end{equation}
Next, we calculate the energy density ratio using the magnetic field from the simulation. 
\begin{align}
    \frac{\varepsilon_{\rm CR, sim}}{\varepsilon_{B, {\rm sim}}} &= \frac{C_{\rm sim}I_\nu/B_{ \perp,~\rm sim}^{\alpha_\nu+1}}{B_{\rm sim}^2/8\pi}\\
    &=\frac{8\pi C_{\rm sim} I_\nu c_{4,\rm sim}^{-\alpha_\nu-1}}{B_{\rm sim}^{\alpha_\nu+3}}.
    \label{eq:resulting_term}
\end{align}
We then multiply Equation~\eqref{eq:resulting_term} with $C_{\rm eq}c_{4, \rm eq}^{-\alpha_\nu-1}/C_{\rm eq}c_{4, \rm eq}^{-\alpha_\nu-1}=1$ such that the numerator can have exact same component as Equation~\eqref{eq:B_eq}, and finally obtain the correlation between the $B_{\rm sim}/B_{\rm eq}$ ratio and the $\varepsilon_B/\varepsilon_{\rm CR}$ ratio:
\begin{align}
    \frac{\varepsilon_{\rm CR, sim}}{\varepsilon_{B, {\rm sim}}} &= \left(\frac{B_{\rm eq}}{B_{\rm sim}}\right)^{\alpha_\nu+3}\left(\frac{c_{\rm 4,sim}}{c_{\rm 4,eq}}\right)^{-\alpha_\nu-1}\frac{C_{\rm sim}}{C_{\rm eq}}
\end{align}
or
\begin{align}
    \log\frac{B_{\rm sim}}{B_{\rm eq}} =&\frac{1}{\alpha_\nu+3}\log\left(\frac{\varepsilon_{B, {\rm sim}}}{\varepsilon_{\rm CR, sim}}\right)\nonumber \\
    &-\frac{\alpha_\nu+1}{\alpha_\nu+3}\log\left(\frac{c_{\rm 4,sim}}{c_{\rm 4, eq}}\right)+\frac{1}{\alpha_\nu+3}\log\left(\frac{C_{\rm sim}}{C_{\rm eq}}\right).
    \label{eq:log-relation-apx}
\end{align}
Here, $c_{\rm 4,sim}$ is the ratio between $B_{\rm perp}$ and $B$ in the simulation. $C_{\rm sim}$ is the proportionality constant with the real $K_0$ and $l$ in the simulation (see also the discussion in Appendix~\ref{apx:ecr_ratio}). A discussion of the offset terms (the second line of Equation~\ref{eq:log-relation-apx}) can be found in Appendix~\ref{apx:offset}. 

\section{Mean deviation of simulated vs.\ equipartition magnetic fields}
\subsection{Offset term in cell-by-cell comparison (Fig.~\ref{fig:cell_by_cell_comparison})}\label{apx:ecr_ratio}
When deriving the equipartition formula, \bk make assumptions that are different from those made in our steady-state modeling. This leads to a deviation between $\varepsilon_{\rm CR,eq}$, calculated with Equation~\eqref{eq:ecr}, and the CR energy density given in the simulation. For example, \bk assume that CR protons and electrons share the same observed spectral index ($\alpha_\mathrm{p}=\alpha_\mathrm{e}$) and hence a constant electron-to-proton ratio. However, in our steady-state post-processing, we assume the CR protons and electrons shares the same injected spectral index, and their observed spectral index is determined by the escape and cooling process. Therefore, the resulting spectrum will not necessarily follow Equation~(A13) in \bk. 

\begin{figure}
    \centering
    \includegraphics[width=\linewidth]{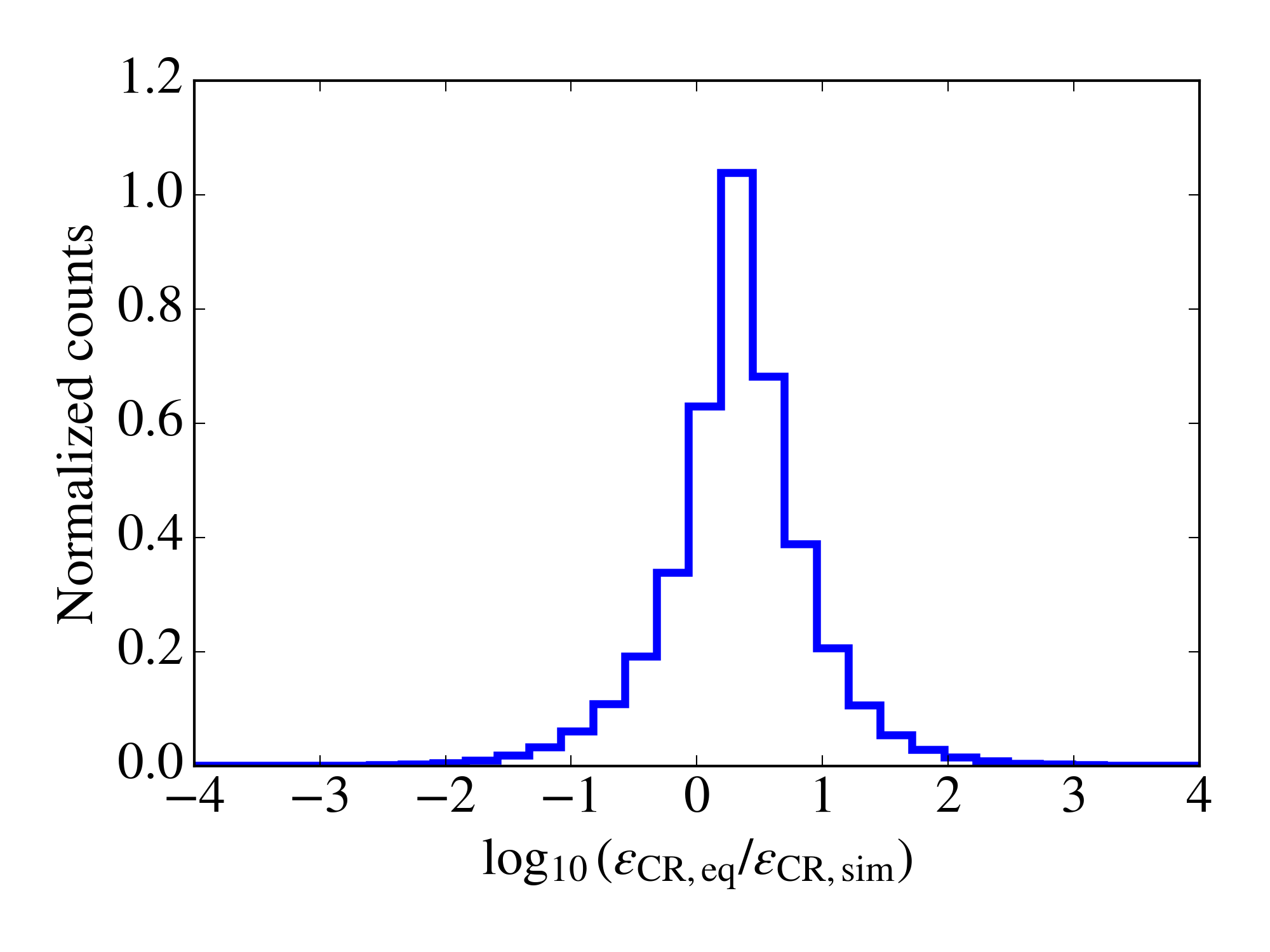}
    \caption{Ratio of the CR energy density calculated by the \bk model ($\varepsilon_{\rm CR, eq}$) and the CR energy density given in the simulation ($\varepsilon_{\rm CR, sim}$). The log-ratio has a mean of 0.32 and a standard deviation of 0.584. }
    \label{fig:ecr_ratio}
\end{figure}

We quantify the difference between the equipartition CR energy density ($\varepsilon_{\rm CR,eq}$, calculated from Equation~\ref{eq:ecr}) and the CR energy density given in the simulation ($\varepsilon_{\rm CR, sim}$) by showing the distribution of the logarithm of their ratio in Fig.~\ref{fig:ecr_ratio}. The mean value of their ratio is 0.32, which means $\varepsilon_{\rm CR,eq}\approx\varepsilon_{\rm CR,sim}\cdot10^{0.32}$. With the slope of the correlation obtained from Fig.~\ref{fig:cell_by_cell_comparison}, we get
\begin{align}
\log \frac{B_{\rm sim}}{B_{\rm eq}} &= 0.29\log\frac{\varepsilon_B}{\varepsilon_{\rm CR, eq}}\\
&=0.29\log\frac{\varepsilon_B}{\varepsilon_{\rm CR,sim\cdot 10^{0.32}}}\\
&=0.29\log\frac{\varepsilon_B}{\varepsilon_{\rm CR,sim}}-0.092.
\end{align}
The offset of 0.092 is very close to the fit in  Fig.~\ref{fig:cell_by_cell_comparison}, where we find an offset of 0.096. This demonstrates that the offset in Fig.~\ref{fig:cell_by_cell_comparison} is largely explained by the erroneous assumption of \bk of an identical CR electron and proton index and constant proton-to-electron ratio for different particle energies.

\subsection{Offset term in face-on maps (Fig.~\ref{fig:map_faceon})}\label{apx:offset}
In this section, we estimate the typical contribution of the second and third terms (the ``offset'' terms) in Equation~\eqref{eq:log-relation} to provide a possible explanation for the offset shown in Fig.~\ref{fig:map_faceon}. This analysis differs from that presented in Appendix~\ref{apx:ecr_ratio}. In Fig.~\ref{fig:cell_by_cell_comparison}, we consider only the magnetic field component perpendicular to the line of sight and perform a cell-by-cell analysis. In contrast, here we evaluate the magnitude of the offset terms as a function of height in order to assess how they influence the projected map.

\begin{figure}
    \centering
    \includegraphics[width=\linewidth]{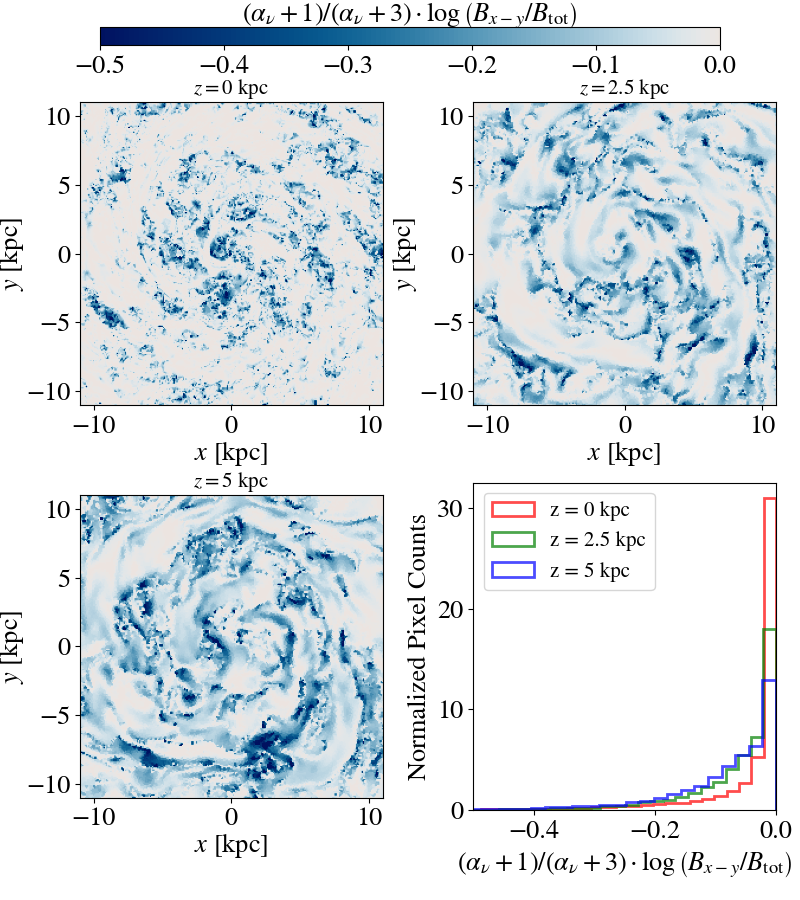}
    \caption{
    Logarithmic ratio of the magnetic field that is parallel to the disk ($B_{\rm sim, x-y}$) and the total magnetic field ($B_{\rm sim, total}$) in a slice at $z=0$, 2.5, and 5 kpc above the disk. With the $(\alpha_\nu+1)/(\alpha_\nu+3)$ pre-factor, the quantity plotted here is equivalent to the second term in the right-hand side of Equation~\eqref{eq:log-relation}, where $c_{\rm 4, eq}=1$, $c_{\rm 4, sim}=B_{\rm sim, x-y}/B_{\rm sim, total}$, and $\alpha_\nu$ is the spectral index slope in each gas cell.  
    The lower right panel shows the normalized distribution of the pixel value in each panel.
    We find that the term $(\alpha_\nu + 1)/(\alpha_\nu + 3) \cdot \log(B_{\rm sim,\ x-y} / B_{\rm sim,\ total})$ peaks near zero and shifts to lower values with increasing height above the disk, indicating that its contribution to the right hand side of Equation~\eqref{eq:log-relation} is small.
    }
    \label{fig:B_reg_perp}
\end{figure}

We begin by plotting the quantity $(\alpha_\nu + 1)/(\alpha_\nu + 3) \cdot \log(B_{\rm sim,\ x\text{-}y} / B_{\rm sim,\ total})$ at different heights above the disk, as shown in Fig.~\ref{fig:B_reg_perp}. In a strictly face-on case, we have $c_{\rm 4, eq} = 1$ and $c_{\rm 4, sim} = B_{\rm sim,\ x\text{-}y} / B_{\rm sim,\ total}$. We find that $B_{\rm sim,\ x\text{-}y} / B_{\rm sim,\ total}$ peaks near unity within the disk and gradually decreases with increasing height. Consequently, the term $(\alpha_\nu + 1)/(\alpha_\nu + 3) \cdot \log(B_{\rm sim,\ x\text{-}y} / B_{\rm sim,\ total})$ follows the same trend, decreasing with height. At $z = 0$, the median value of this term is $-0.01$, indicating that it is not a dominant contributor to the offset in the face-on map.

\begin{figure}
    \centering
    \includegraphics[width=\linewidth]{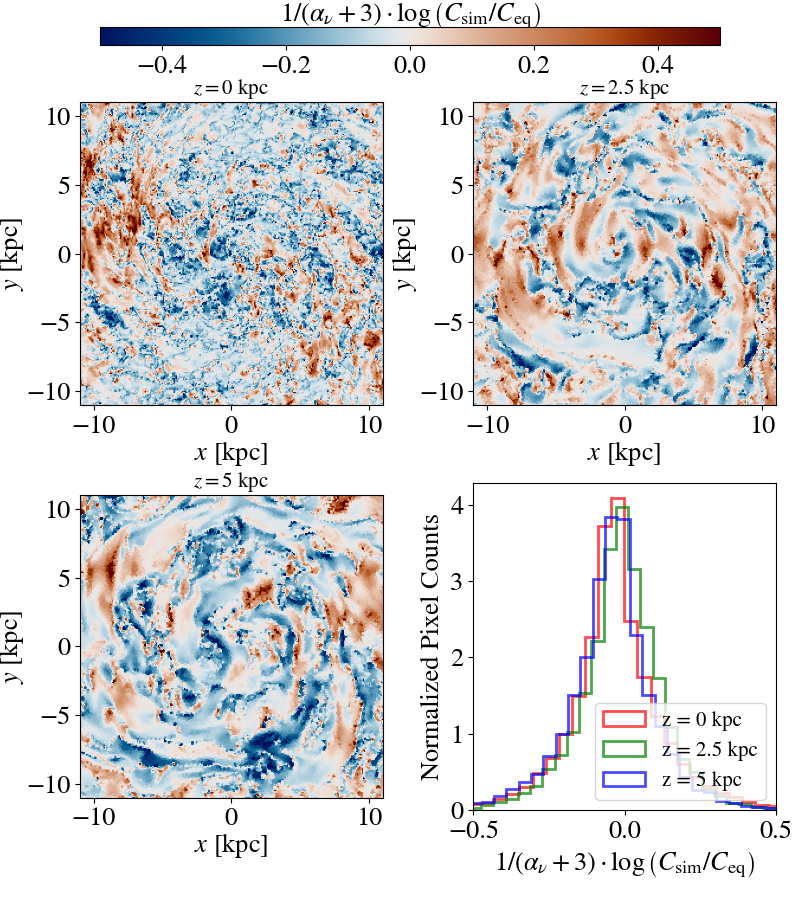}
    \caption{Logarithmic ratio of the proportionality constants in the equipartition formula derivation, $C_{\rm eq}=f(\alpha_\mathrm{e})(K_0+1)/l$, and in the simulation, $C_{\rm sim}=\varepsilon_{\rm CR}/(I_\nu/B_\perp^{\alpha_\nu+1})$, scaled by the pre-factor $1/(\alpha_\nu + 3)$ to estimate the contribution of the third term in Equation~\eqref{eq:log-relation}. The lower-right panel presents the normalized distribution of pixel values across all panels. This term contributes only marginally, as the distribution peaks around zero.}
    \label{fig:C_ratio}
\end{figure}

We then examine the contribution from the third term, $1/(\alpha_\nu + 3) \cdot \log(C_{\rm sim} / C_{\rm eq})$, as shown in Fig.~\ref{fig:C_ratio}. As defined in Equation~\eqref{eq:ecr}, $C$ is the proportionality constant relating the CR energy density to $I_\nu / B_\perp^{\alpha_\nu + 1}$. In \bk, this constant depends on $\alpha_\nu$, $K_0$, $l$, and the assumption that CR electrons and protons share the same spectral slope. In our simulation, we define it effectively as $C_{\rm sim} = \varepsilon_{\rm CR, sim} / (I_\nu / B_\perp^{\alpha_\nu + 1})$. We plot the quantity $1/(\alpha_\nu + 3) \cdot \log(C_{\rm sim} / C_{\rm eq})$, including the pre-factor, for slices at various heights above the disk in Fig.~\ref{fig:C_ratio}, with the corresponding normalized distributions shown in the lower-right panel. The distribution peaks near zero, with a median value of $-0.04$, slightly larger in magnitude than the contribution from the second term. This suggests that the third term plays a somewhat more significant role than the second term in the face-on map. 

This value does not exactly match the offset shown in Fig.~\ref{fig:map_faceon}, for the following reason. In the fit, we compare $\langle\log(\varepsilon_B / \varepsilon_{\rm CR})\rangle$ with $\log(\langle B_{\rm sim} \rangle / B_{\rm eq})$ to reflect quantities accessible in observations. However, $\log(\langle B_{\rm sim} \rangle / B_{\rm eq})$ differs from $\langle\log(B_{\rm sim} / B_{\rm eq})\rangle$, which corresponds to directly averaging both sides of Equation~\eqref{eq:log-relation}. The latter can only be computed in simulations via cell-by-cell comparisons, as $B_{\rm sim}$ is not directly observable. Therefore, the small slope and offset in the fitting lines shown in Fig.~\ref{fig:map_faceon} and Fig.~\ref{fig:map_edgeon} are non-trivial, yet remain consistent with the predictions of Equation~\eqref{eq:log-relation}. 

So far, we only discussed the contributions of the offset terms to the mean. However, the fact that both terms show a non-vanishing distribution around their mean values implies that both terms contribute to the scatter of the field strength ratio $B_{\rm sim} / B_{\rm eq}$, with a dominant contribution of the last term in Equation~\eqref{eq:log-relation}. 

\section{On which scale does the equipartition formula work?}\label{apx:scale}
\begin{figure}
    \centering
    \includegraphics[width=\linewidth]{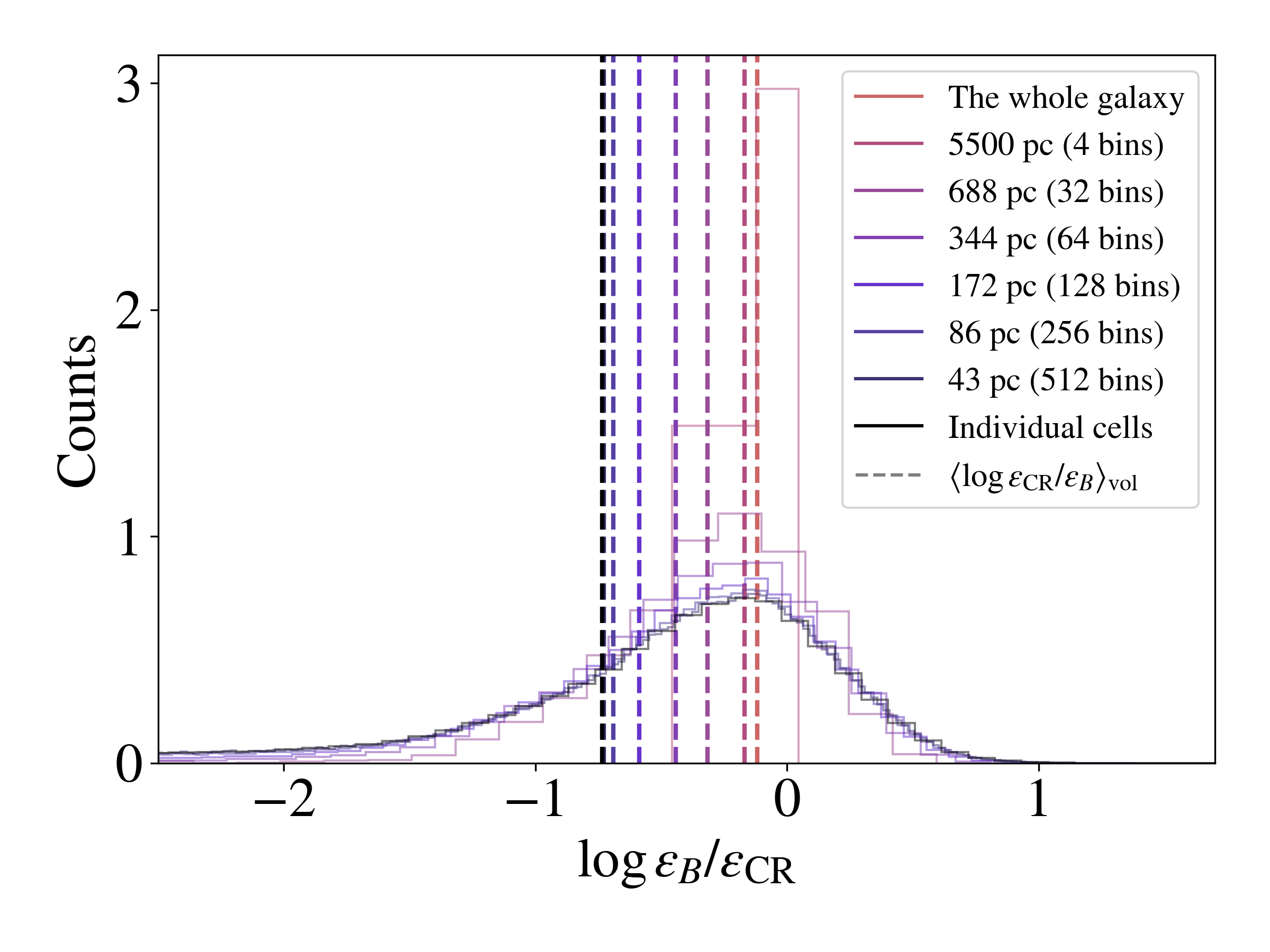}
    \caption{Average energy density ratio distributions on different spatial scales. Different colors represent varying physical averaging scales corresponding to different numbers of bins. Dashed line represents the mean value of the distribution of the corresponding color. 
    Both the distributions and their mean values are volume-weighted. The gradual shift in the mean value from lower to higher values of $\varepsilon_B/\varepsilon_{\rm CR}$ indicates that equipartition holds better on progressively larger scales while transiting toward CR energy dominance at smaller scales.
    }
    \label{fig:eb_ecr_ratio}
\end{figure}

Previous studies have shown that energy equipartition does not hold on scales smaller than $\sim1$~kpc by comparing the fluctuation of synchrotron intensity and the magnetic field \citep{stepanov_observational_2014, seta_revisiting_2019}. In Fig.~\ref{fig:eb_ecr_ratio}, we examine this argument with our simulation data. In this analysis, we consider all gas cells within the disk ($|x|<11$~kpc, $|y|<11$~kpc, $|z|<0.5$~kpc). 
We partition the domain of interest into cubes of different volumes, implying a different number of bins. In each cube, we calculate the average energy density of both CR and the magnetic field, and plot the distribution of the ratio (weighted by the volume in each cube) as a histogram in the background of Fig.~\ref{fig:eb_ecr_ratio}, overlaying their mean value as the vertical dash line. The $z$ direction binning is chosen such that they have similar spatial resolution with the $x$ and $y$ direction. For example, in the case of 4 bins, we do not apply any cuts in $z$ direction; in the case of 32 bins, the $z$ direction is partitioned into 2 bins (instead of 32 bins), and so on.

We find that the CR and magnetic energy density ratio is approximately in equipartition when considering the entire galaxy, but gradually shifts toward CR dominance at the level of individual cells. This agrees with previous studies \citep{stepanov_observational_2014, seta_revisiting_2019} and shows that equipartition does not hold on small scales. We explain this trend as follows. In the lower right panel of Fig.~\ref{fig:map_faceon}, the volume-weighted ratio $\langle\log \varepsilon_B/\varepsilon_{\rm CR}\rangle_{\rm vol}$ is higher in the spiral arms and lower in the inter-arm regions because CR energy density is more uniformly distributed than magnetic energy density and magnetic energy density are stronger in the spiral arms. 
Specifically, CR energy usually dominates in more diffusive regions, such as the outer region of the galaxy. Therefore, the volume-weighted average tends to emphasize CR-dominant regions, and bias the ratio $\langle\log \varepsilon_B/\varepsilon_{\rm CR}\rangle_{\rm vol}$ toward smaller values. On the other hand, on increasingly larger scales, extreme magnetic energy densities are averaged with more extended low-field regions, smoothing out local fluctuations and revealing the galaxy's underlying equipartition on large scales, as both magnetic and CR energy densities are linked to star formation activity.

\begin{figure}
    \centering
    \includegraphics[width=\linewidth]{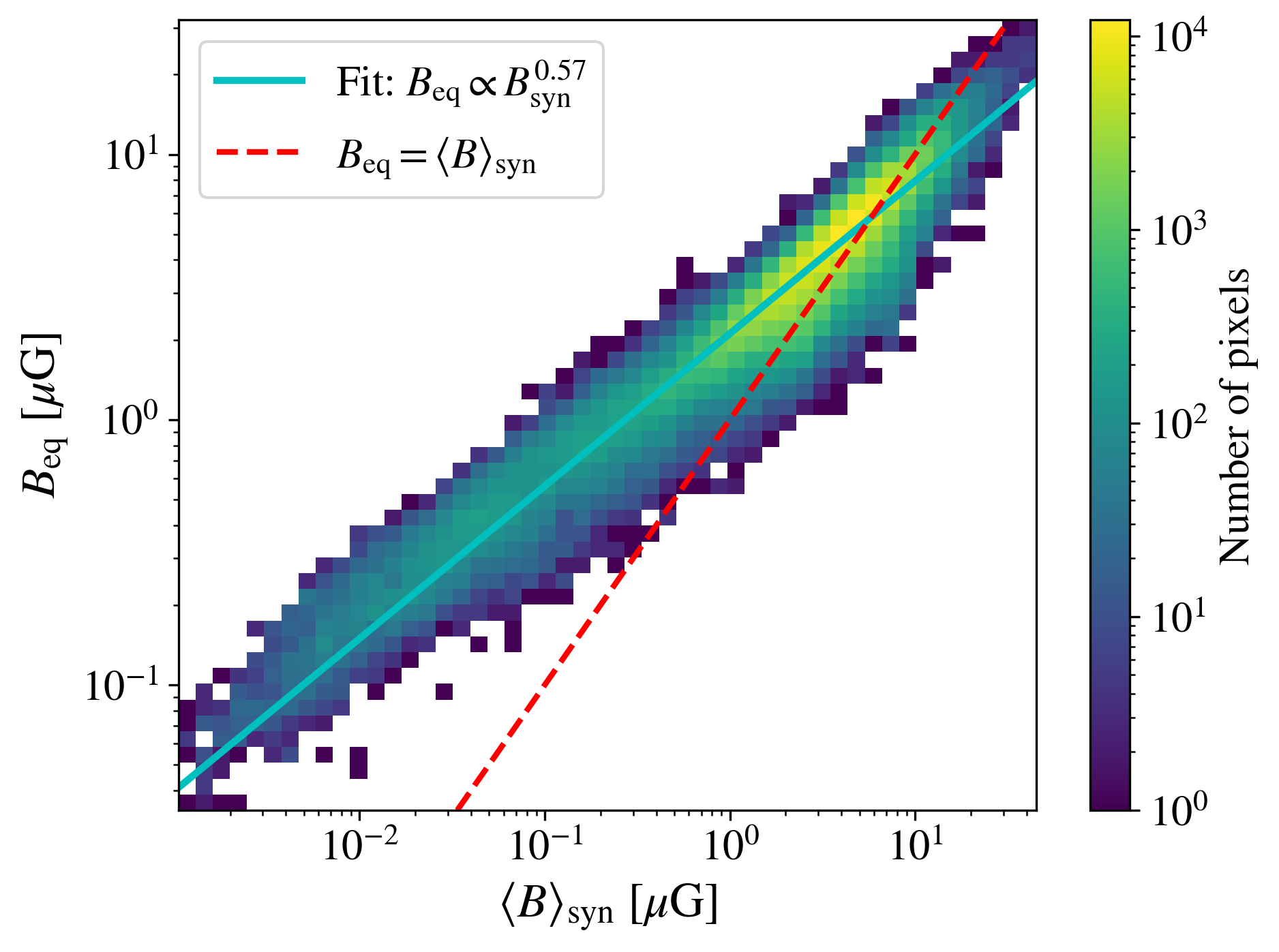}
    \caption{2D distribution between synchrotron-weighted and the equipartition magnetic field. The pixel-by-pixel correlation shows that the equipartition magnetic field traces the synchrotron-weighted field on the smallest scale of our projection map, which is 22~kpc/512~pixels $\sim$ 42~pc. 
    }
    \label{fig:lucia}
\end{figure}

However, as our results demonstrate, the key issue is not whether equipartition is strictly valid, but whether the equipartition formula can still provide a reliable estimate of the synchrotron-weighted magnetic field. In Fig.~\ref{fig:lucia}, we plot a pixel-by-pixel comparison between the projection map of the synchrotron-weighted magnetic field and the equipartition magnetic field in face on view, enabling a direct comparison with \citet{linzer_estimation_2025}. We find that most of the pixels (colored yellow) are located near the $B_{\rm eq}=\langle B\rangle_{\rm syn}$ line (red-dash line). This suggests that statistically the formula remains effective in practice for face-on inclination, and that the equipartition method can be a useful diagnostic tool at sub-kiloparsec scales, even when its core assumptions are not fully satisfied.

In \citet{linzer_estimation_2025}, the correlation between $B_{\rm eq}$ and $B_{\rm syn}$ is explained by assuming the CR energy density is uniform in their domain of interest. With $I_\nu \propto B_{\rm sim}^{\alpha_\nu+1}\varepsilon_{\rm CR}$ and $\varepsilon_{\rm CR}$ being constant, $B_{\rm eq}\propto I_\nu^{1/(\alpha_\nu+3)}\propto B_{\rm sim}^{(\alpha_\nu+1)/(\alpha_\nu+3)}$. However, in our simulation, we found that $\varepsilon_{\rm CR}$ is roughly\footnote{Under the assumption used in BK05, when CR protons and electrons share the same spectral index, $j_\nu\propto B_{\rm sim}^{\alpha_\nu+1}\varepsilon_{\rm CR}$. Combined with the relation $j_\nu\propto B_{\rm sim}^{\alpha_\nu+1}/(\varepsilon_B+\varepsilon_{\rm ph})$ \citep{pfrommer_simulating_2008}, we have $\varepsilon_{\rm CR}\propto1/(\varepsilon_B+\varepsilon_{\rm ph})$. This means that CR energy density is only constant when the photon energy density $\varepsilon_{\rm ph}$ is much larger than the magnetic energy density. When considering a purely advective CR transport model with adiabatic compression ($\varepsilon_{\rm CR}\propto \rho^{4/3}$) and flux freezing of the magnetic field ($\varepsilon_{B}\propto \rho^{2/3}$), $\varepsilon_{\rm CR}\propto B^2$. In conclusion, CR energy density can vary from $\varepsilon_{\rm CR}\propto B^{-2}$ to $\varepsilon_{\rm CR}\propto B^2$, while diffusion and streaming will further modify this correlation \citep{2025arXiv251016125T}.} $\propto B^{0.42}$, and hence the relation becomes $B_{\rm eq}\propto B_{\rm sim}^{(\alpha_\nu+1.42)/(\alpha_\nu+3)}$. With the same range of $\alpha_\nu=0.5-1.15$, the exponent becomes 0.55-0.62, consistent with the slope obtained in Fig.~\ref{fig:lucia}. 

\end{appendix}

\end{document}